\documentclass[lettersize,journal]{IEEEtran}
\newcommand{\RNum}[1]{\uppercase\expandafter{\romannumeral #1\relax}}
\usepackage{mathrsfs}
\usepackage{amssymb}
\usepackage{diagbox}
\usepackage{cite}
\usepackage[T1]{fontenc}
\usepackage{graphicx}
\usepackage{CJKutf8}
\usepackage{xspace}
\usepackage{makecell}
\usepackage{psfrag}
\usepackage{url}
\usepackage{stfloats}
\usepackage{amsmath}
\usepackage{array}
\usepackage{float}
\usepackage{multirow} 
\usepackage{hyperref}
\usepackage{amsthm}
\usepackage{color}
\usepackage{multicol}
\usepackage{stfloats}
\usepackage{enumerate}
\usepackage[utf8]{inputenc} % allow utf-8 input
\usepackage[T1]{fontenc}    % use 8-bit T1 fonts
\usepackage{hyperref}       % hyperlinks
\usepackage{url}            % simple URL typesetting
\usepackage{booktabs}       % professional-quality tables
\usepackage{amsfonts}       % blackboard math symbols
\usepackage{nicefrac}       % compact symbols for 1/2, etc.
\usepackage{microtype}      % microtypography
\usepackage{color}
\usepackage{xcolor}
\usepackage{graphicx}
\usepackage{subcaption}
\usepackage{multirow}
\usepackage{flushend}
\usepackage{soul}
\usepackage{mathrsfs}
\usepackage{bbm}
\usepackage{amssymb}

\usepackage{psfrag,calc,url,bm}

\usepackage{cite}

\usepackage{graphicx}

\usepackage{psfrag}

\usepackage{url}

\usepackage{stfloats}

\usepackage{amsmath}

\usepackage{float}

\usepackage{algorithmic}

\usepackage[ruled,linesnumbered,vlined]{algorithm2e}

\usepackage{color}

\usepackage{boxedminipage}

\usepackage{amsthm}

\usepackage{multirow}

\usepackage{setspace}

\usepackage{array}
\usepackage{wrapfig}
\usepackage{bbm}
\usepackage{mathtools}
\usepackage{dsfont}
\colorlet{soulgray}{lightgray!30}
\sethlcolor{soulgray}
\usepackage{booktabs}  %±í¸ñ
\newtheorem{remark}{Remark}

\newtheorem{definition}{Definition}

\allowdisplaybreaks[4]
\usepackage[ruled]{algorithm2e}
\usepackage{algorithmic} %format of the algorithm
\ifCLASSINFOpdf
\else
\fi
\begin{document}
\begin{CJK}{UTF8}{gbsn}
% \title{Provisioning Zero-Trust Services via Hierarchical Micro-Segmentations: An LLM-Enhanced Graph Diffusion Approach}
\title{Hierarchical Micro-Segmentations for Zero-Trust Services via Large Language Model (LLM)-enhanced Graph Diffusion}

%AI-empowered Adaptive Coverage Enhancement in 6G: Opportunities and Challenges

%\author{\IEEEauthorblockN{Michael Shell}
%\IEEEauthorblockA{School of Electrical and\\Computer Engineering\\
%Georgia Institute of Technology\\
%Atlanta, Georgia 30332--0250\\
%Email: http://www.michaelshell.org/contact.html}
%\and
%\IEEEauthorblockN{Homer Simpson}
%\IEEEauthorblockA{Twentieth Century Fox\\
%\and
%\IEEEauthorblockN{James Kirk\\ and Montgomery Scott}
%\IEEEauthorblockA{Starfleet Academy\\
%San Francisco, California 96678-2391\\
%\author{Ruichen Zhang$^*$ Ke Xiong$^*$, Wei Guo$^*$ \\
%\small
%$^*$School of Computer and Information Technology, Beijing Jiaotong University, Beijing 100044, P. R. China, \\
%Beijing 100044, China\\
%E-mail:18120449@bjtu.edu.cn}
	\author{Yinqiu Liu$^{*}$, Guangyuan Liu$^{*}$, Hongyang Du, Dusit Niyato,~\textit{Fellow, IEEE}, Jiawen Kang, \\Zehui Xiong, Dong In Kim,~\textit{Fellow, IEEE}, and Xuemin (Sherman) Shen,~\textit{Fellow, IEEE} 
 \thanks{Y. Liu, G. Liu, H. Du, and D. Niyato are with the College of Computing and Data Science, Nanyang Technological University, Singapore (e-mail: yinqiu001@e.ntu.edu.sg, liug0022@e.ntu.edu.sg, hongyang001@e.ntu.edu.sg, and dniyato@ntu.edu.sg).}

\thanks{J. Kang is with the School of Automation, Guangdong University of Technology, China (e-mail: kavinkang@gdut.edu.cn).}

\thanks{Z. Xiong is with the Pillar of Information Systems Technology and Design, Singapore University of Technology and Design, Singapore (e-mail: zehui\_xiong@sutd.edu.sg).}

\thanks{D. I. Kim is with the Department of Electrical and Computer Engineering, Sungkyunkwan University, South Korea (email: dongin@skku.edu).}

\thanks{X. Shen is with the Department of Electrical and Computer Engineering, University of Waterloo, Canada (e-mail: sshen@uwaterloo.ca).}

%\thanks{Y. Liu and G. Liu contributed equally to the work.}
  \vspace{-1.55em}}

\maketitle

\begin{abstract}
In the rapidly evolving Next-Generation Networking (NGN) era, the adoption of zero-trust architectures has become increasingly crucial to protect security. 
However, provisioning zero-trust services in NGNs poses significant challenges, primarily due to the environmental complexity and dynamics. 
Motivated by these challenges, this paper explores efficient zero-trust service provisioning using hierarchical micro-segmentations. 
Specifically, we model zero-trust networks via hierarchical graphs, thereby jointly considering the resource- and trust-level features to optimize service efficiency.
We organize such zero-trust networks through micro-segmentations, which support granular zero-trust policies efficiently.
To generate the optimal micro-segmentation, we present the Large Language Model-Enhanced Graph Diffusion (LEGD) algorithm, which leverages the diffusion process to realize a high-quality generation paradigm. 
Additionally, we utilize policy boosting and Large Language Models (LLM) to enable LEGD to optimize the generation policy and understand complicated graphical features.
Moreover, realizing the unique trustworthiness updates or service upgrades in zero-trust NGN, we further present LEGD-Adaptive Maintenance (LEGD-AM), providing an adaptive way to perform task-oriented fine-tuning on LEGD.
Extensive experiments demonstrate that the proposed LEGD achieves 90\% higher efficiency in provisioning services compared with other baselines. Moreover, the LEGD-AM can reduce the service outage time by over 50\%.
\end{abstract}


\vspace{-0.1cm}
\section{Introduction}
% \subsection{Background}
In cybersecurity, conventional trust models operate on the principle that once one entity is authenticated, it can permanently access network resources \cite{10052642}. 
This implicit trust creates significant vulnerabilities, as it allows attackers who breach the perimeter from inside to move laterally, thus damaging the entire network \cite{10052642}. 
Zero-trust, however, represents a fundamental shift from this paradigm. 
Specifically, it follows the \textit{never trust, always verify} principle, ensuring that every access request is authenticated, authorized, and continuously validated \cite{10052642}. 
By strictly enforcing least-privilege access, both the risk of insider threats and outsider attacks can be alleviated \cite{10437186}. 
For Next-Generation Networking (NGN) \cite{10437338}, the demand for zero-trust is projected to be higher than ever, driven by the increasing volume of data, devices, and applications.

% \subsection{Motivation}
Recent academic research has made significant strides in advancing zero-trust, especially in two aspects. 
First, as the prerequisite for applying zero-trust policies, abstract trustworthiness levels/relationships should be represented quantitatively.
Most existing proposals \cite{10488084, 10330693, 10292665, 10138335} evaluate trustworthiness based on reputation, which dynamically adjusts the trust score of each entity according to its historical behaviors.
%In addition, tag \cite{tag} and social theory \cite{10091249} can also be adopted.
With quantitative trustworthiness, zero-trust authentications/access control for various scenarios can be developed. 
For instance, Feng et al. \cite{10330693} presented the multi-layer access control for zero-trust cyber-physical systems. 
Ge et al. \cite{10292665} proposed interdependent trust evaluation and authentication policies using dynamic game models.
Additionally, Cheng et al. \cite{10138335} leveraged blockchain to design a decentralized voting-based authentication protocol for zero-trust 5G communications. 
%Additionally, Abuhasel et al. \cite{10287925} proposed a zero-trust architecture for Industry 5.0, focusing on secure authentication to facilitate human-machine information sharing.
%The integration of zero-trust with other emerging techniques, such as blockchain and federated learning \cite{9732238}, has also been widely researched.
%For instance, Liu et al. \cite{9732238} investigated blockchain-based decentralized zero-trust, utilizing blockchain's immutability to create tamper-proof verification systems.
%Furthermore, Hussain et al. \cite{10495909} applied federated learning to enable zero-trust, allowing privacy-preserved model training.
Apart from academia, various industrial zero-trust solutions have been presented, such as Entra \cite{Entra} of Microsoft and MaaS360 \cite{MaaS360} of IBM.
Despite such progress, one critical problem is rarely studied, i.e., the \textit{service provisioning \cite{6933929} for next-generation zero-trust networks}.
We notice that two daunting challenges exist to solve this issue
\begin{itemize}
    \item \textbf{Challenge 1}: Zero-trust relies on granular authentication and verification policies \cite{10052642}. Despite enhancing security, it complicates service provisioning, especially for NGN services involving multiple steps handled by different entities. Additionally, zero-trust policies for different applications and network sectors vary significantly \cite{Micro2}. For instance, medical systems require stricter, multi-layer authentication policies due to the extensive privacy concerns, unlike social blogs \cite{7990130}. Consequently, directly submitting service requests to zero-trust networks often results in frequent access denials, adversely affecting user experience and resource efficiency.
    \item \textbf{Challenge 2}: Zero-trust networks exhibit great dynamics due to the necessity for continuous trustworthiness maintenance \cite{9773102}. For instance, a widely trusted entity with a high reputation may be removed in the next period due to a single malicious behavior \cite{8685209}. Consequently, the service provisioning strategies defined at the initial stage may quickly become outdated, leading to significant performance degradation.
\end{itemize}

% \subsection{Our Work and Contribution}
In this paper, we present efficient service provisioning for zero-trust NGNs through micro-segmentations \cite{Micro2}. This approach offers two main advantages: 1) it enables granular and independent zero-trust policies, ensuring that if one micro-segmentation is compromised, others can remain unaffected, and 2) it enhances flexible resource scheduling by specializing each micro-segmentation in a specific service type \cite{NIST}. In the NGN era, each service is accomplished by a Service Function Chain (SFC) \cite{10298057} with multiple steps, posing the challenge of generating optimal micro-segmentations to maximize service efficiency. To this end, we model zero-trust networks using a hierarchical graph that integrates physical resources and trustworthiness relationships. Then, we present the Large Language Model-Enhanced Graph Diffusion (LEGD) algorithm, leveraging state-of-the-art diffusion-based generative architecture \cite{diffusion} for controllable micro-segmentation generation. Inspired by human feedback mechanisms \cite{RLHF}, we incorporate an LLM-empowered agent, which generates and activates heuristic filters to improve LEGD's efficiency. Finally, we propose an adaptive micro-segmentation maintenance mechanism that promptly updates micro-segmentations to adapt to continuous trustworthiness updates and service upgrades in zero-trust NGNs. Our main contributions are summarized as follows.
\begin{itemize} 
\item \textbf{Zero-trust Service Provisioning Framework}: \textit{To the best of our knowledge, this is the first work that systematically studies efficient service provisioning for zero-trust NGN.} We propose a novel framework that organizes the zero-trust network via micro-segmentations and provisions services by SFCs. Using graph theory, we model zero-trust networks through a hierarchical graph, which jointly considers physical and trust-level features to optimize the service utility.  
\item \textbf{LEGD for Micro-segmentation Generation}: We present the LEGD algorithm for controllable micro-segmentation generation. Leveraging diffusion architecture, LEGD exhibits excellent exploration capability via a denoising process. We adopt policy boosting to optimize the generation policy, allowing LEGD to reinforce itself through interacting with zero-trust networks. We also propose an LLM-empowered agent to provide human-like perceptions of the graphical network environment, thus activating heuristic filters to improve LEGD's efficiency. 
\item \textbf{Adaptive Micro-segmentation Maintenance}: To adapt to the dynamic zero-trust NGN environments, we propose an adaptive micro-segmentation maintenance algorithm named LEGD-Adaptive Maintenance (LEGD-AM). By fine-tuning well-trained LEGD models, LEGD-AM can respond to trustworthiness updates and service upgrades promptly. Moreover, adaptive masks and reward engineering are developed to ensure LEGD-AM aligns with updated environments and tasks.
% \item \textbf{Experimental Results}: We perform extensive experiments to validate our proposals. Specifically, our LEGD algorithm archives over 90\% performance gain in generating micro-segmentations and provisioning services, compared with the conventional method. Moreover, the LEGD-AM can effectively adapt to both network topology and service changes, saving over 50\% of the time than the conventional method.
\end{itemize}

% \subsection{Organization}
The remainder of this paper is organized as follows. 
Section II reviews related works. 
The system models are described in Section III. 
Then, in Section IV, we elaborate on the design of the LEGD algorithm. 
LEGD-AM is described in Section V. 
The numerical results are discussed in Section VI. 
Finally, Section VII concludes the paper.

\section{Related Work and Motivation}
\subsection{Zero-trust Network Architecture}
Unlike the static trust model, zero-trust follows the principle of ``never trust, always verify'' by performing per-session authentication and continuous identification \cite{10052642}. Trustworthiness evaluations, such as those by reputation \cite{8685209}, social theory \cite{10091249}, and tags \cite{tag}, accurately reflect the trust level of each entity. Rational trustworthiness values enable granular zero-trust policies in various scenarios. For example, zero-trust mechanisms for cyber-physical systems and the Metaverse were presented in \cite{10330693} and \cite{10138335}, respectively. Additionally, Chen et al. \cite{9273056} designed dynamic access control for 5G smart healthcare. Researchers have also optimized zero-trust architectures and algorithms for network authentication \cite{10437443}. Despite these advancements, service provisioning in zero-trust networks is still under-researched. The approach in \cite{10310190} used service function chains for scheduling but only supported heuristic device selection by trustworthiness without considering the physical and trust-level network environments. Motivated by this, we intend to provide a learning-based approach for optimal service provisioning. 

\subsection{Micro-segmentation}
Micro-segmentation is regarded as the fundamental manner for organizing zero-trust NGNs, enabling granular network isolation to limit the lateral movement of threats \cite{NIST, Auze}. 
Early micro-segmentations are generated in static ways \cite{9773102}, which are often rigid and lack the flexibility facing dynamic environments. 
Contemporary approaches, such as illumio \cite{illumio} and \cite{Micro2}, leveraged Software-Defined Networking (SDN) to achieve programmable micro-segmentation, enabling real-time policy enforcement to defend against intelligent attacks. 
Nonetheless, challenges still remain in ensuring the generated micro-segmentation can maximize user experience and service efficiency, especially in large heterogeneous NGNs. 
To this end, Yousefi-Azar et al. \cite{DRLMicro} explored this issue by fine-tuning segmentation strategies via unsupervised learning. 
In this paper, we define multiple performance indicators for evaluating zero-trust services.
Moreover, we abstract micro-segmenting as a generation problem and present the LEGD algorithm based on the state-of-the-art generative model named diffusion \cite{diffusion} to realize controllable micro-segmentation generation.

\subsection{Graph Neural Networks in Networking}
Graphs can model various networking and communication problems like resource allocation, routing optimization, and traffic prediction \cite{9979700}. This is because network topologies naturally form graph structures, enabling efficient representation of complex relationships. Graph Neural Networks (GNNs) have been extensively studied for these problems. For example, Liu et al. \cite{10285729} used graph attention networks to predict connection failures. Peng et al. \cite{10401242} introduced Vertex-GNN and Edge-GNN for resource allocation in wireless networks. Hou et al. \cite{10437541} developed a GNN-based caching scheme for SDN-based networks. Micro-segmentation generation, however, differs from conventional graphical optimizations. First, generation problems are inherently discrete, requiring decisions about the presence/absence of nodes and edges. Moreover, different from discrete optimization problem that searches through the solution space, generation focuses on producing the desired data distributions \cite{diffusion}. To this end, we present the LEGD with the state-of-the-art generative architecture, i.e., diffusion.

\section{System Model: Hierarchical Micro-Segmentations for Zero-trust Services}
\subsection{Network Modeling via Hierarchical Graph}
We consider efficient service provisioning in the zero-trust context.
Without loss of generality, we take AI-Generated Content (AIGC) \cite{10221755}, one of the most emerging applications in the NGN era, as an example to illustrate system design.

\subsubsection{Stakeholder} 
As shown in Fig. \ref{TR}, the system accommodates two types of stakeholders, namely users and service providers.
Users generate AIGC works by requesting service providers.
%PEs act as the coordinator of zero-trust, whose functions are discussed below.
Particularly, considering the multimodality feature of AIGC, we suppose one generation task may contain multiple steps.
Every step should be handled by a specific type of service provider owning the corresponding AIGC models.
For instance, Fig. \ref{TR}-A showcases a three-step AIGC creation chain, in which the user leverages textual descriptions to draw images and then selects the most preferred one to make videos.
Such AI-generated video services have accumulated a market size of more than USD 554.9 million till 2023\footnote{Data available at: https://www.grandviewresearch.com/industry-analysis/ai-video-generator-market-report}.
Accordingly, the SFC should consist of text generation, text-to-image, and image-to-video service providers in sequence.

\subsubsection{Network}
\begin{figure}[tbp!]
  \centering
  \includegraphics[width=0.5\textwidth]{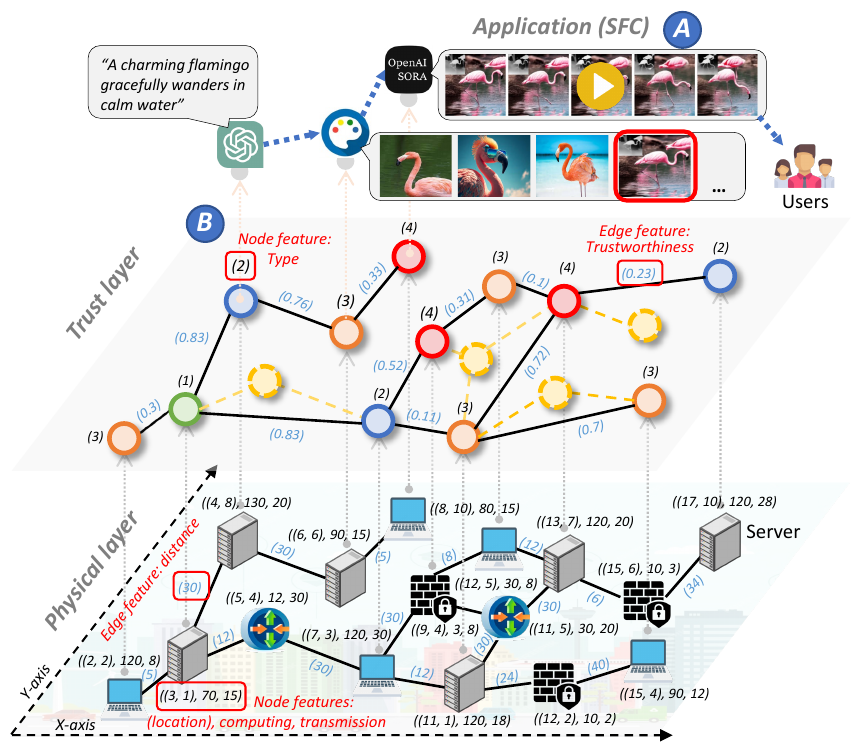} 
  \caption{System model. \textbf{A}: The illustration of a three-step SFC. \textbf{B}: The zero-trust network modeling using a hierarchical graph.} 
  %\vspace{0.2cm}
  \label{TR}
\end{figure}

We model the network topology using a hierarchical graph with physical and trust layers.
As shown in Fig. \ref{TR}-B, the physical layer is represented by a directed graph $\boldsymbol{G}^P(\boldsymbol{V}^P, \boldsymbol{E}^P, \boldsymbol{F}_V^P, \boldsymbol{F}_E^P)$, where all the devices construct node set $\boldsymbol{V}^P$, and all the communication links between each pair of devices form edge set $\boldsymbol{E}^P$.
Accordingly, $\boldsymbol{F}_V^P$ and $\boldsymbol{F}_E^P$ preserve the node- and edge-level features, respectively, i.e.,
\begin{gather}
    \boldsymbol{F}_V^P = \left \{f_{v_i}^P= \left((x_i, y_i), c(v_i), \varpi(v_i)\right)\;|\; \forall v_i \in \boldsymbol{V}^P \right \}, \\
    \boldsymbol{F}_E^P = \left \{f_{e_{i\rightarrow j}}^P= \left(d(v_i, v_j)\right)\;|\; \forall e_{i\rightarrow j} \in \boldsymbol{E}^P \right \},
\end{gather}
where ($x_i$, $y_i$), $c(v_i)$, and $\varpi(v_i)$ indicate the 2D-location, computing, and transmission power of node $v_i$, respectively.
$d(v_i, v_j) = \sqrt{(x_i - x_j)^2 + (y_i -y_j)^2}$ denotes length of edge $e_{i \rightarrow j}$, i.e., the physical distance between nodes $v_i$ and $v_j$.

Unlike conventional one-time identifications, zero-trust requires dynamic trustworthiness management and per-session authentications.
To this end, a trust layer is built atop the physical layer, preserving the network-wide trustworthiness relationship.
Specifically, we leverage an undirected graph $\boldsymbol{G}^R(\boldsymbol{V}^R, \boldsymbol{E}^R, \boldsymbol{F}_V^R, \boldsymbol{F}_E^R)$ to model the trust layer, where $\boldsymbol{V}^R \in \boldsymbol{V}^P$ and $\boldsymbol{E}^R \in \boldsymbol{E}^P$, denoting the participants of zero-trust service provisioning.
Accordingly, the node and edge features $\boldsymbol{F}^R_V$ and $\boldsymbol{F}^R_E$ are defined as
\begin{gather}
    \boldsymbol{F}_V^R = \left \{f_{v_i}^R= \left(s_i\right)\;|\; \forall v_i \in \boldsymbol{V}^R \right \}, \\
    \boldsymbol{F}_E^R = \left \{f_{e_{i\leftrightarrow j}}^R= (\omega_{i \leftrightarrow j})\;|\; \forall e_{i\leftrightarrow j} \in \boldsymbol{E}^R \right \},
\end{gather}
where $s_i$ represents the roles that the node plays in SFC, i.e., type-\{1, 2, $\dots$, $K$\} service provider.
$\omega_{i \leftrightarrow j}$ denotes the mutual trustworthiness between nodes $v_i$ and $v_j$ $\in$ $\boldsymbol{V}^R$, which is defined below.
%Note that mutual trustworthiness is defined by
%\begin{equation}
%    \omega_{i \leftrightarrow j} = \omega_{j \leftrightarrow i} = \mathcal{M}(v_i, v_j) \in [0, 1], \forall v_i, v_j \in \boldsymbol{V}^R.
%\end{equation}
%The higher the $\omega_{i \leftrightarrow j}$ value, the higher the level of trust that is maintained.
%To ensure the adaptability of our proposal, the trustworthiness evaluation scheme $\mathcal{M}(v_i, v_j)$ is regarded as a pluggable module, supporting various well-proven schemes, such as reputation \cite{8685209} and tags \cite{tag}.
Through this hierarchical graph, we can perform joint optimization regarding resources and trustworthiness.
% Next, we discuss the modeling of two layers in detail.
\begin{figure*}[tbp!]
  \centering
  \includegraphics[width=0.95\textwidth]{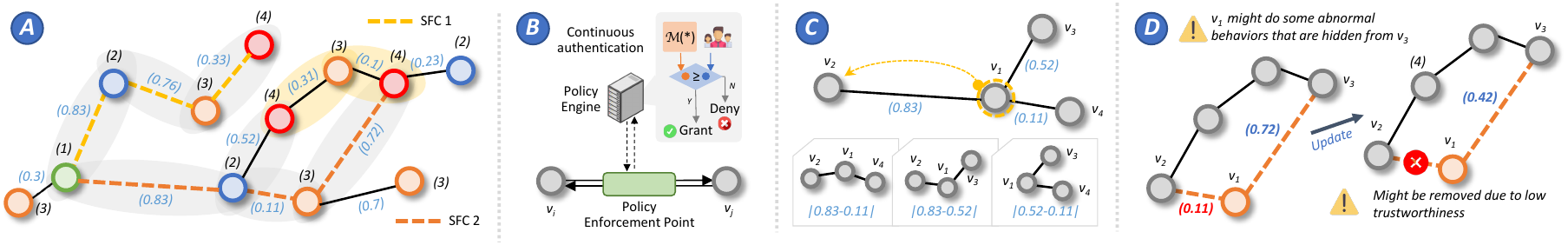} 
  \caption{\textbf{A}: The illustration of two micro-segmentations (marked by gray and yellow, respectively) over the trust layer of Fig. 1. \textbf{B}: The zero-trust policy. \textbf{C}: The calculation of $E_s$ on node $v_1$. \textbf{D}: The example to explain trustworthiness equilibrium.} 
  \vspace{-0.4cm}
  \label{EX}
\end{figure*}

\subsubsection{Zero-Trust Policy and Micro-segmentation}
As shown in Fig. \ref{EX}-A, the entire trust layer is partitioned into multiple isolated micro-segmentations, each of which accommodates multiple SFCs catering to a specific service type.
Within each micro-segmentation, we suppose the NIST standards \cite{NIST} are adopted to implement zero-trust policies (see Fig. \ref{EX}-B).
Specifically, each pair of nodes is monitored by one Policy Enforcement Point (PEP).
All PEPs are coordinated and managed by a centralized Policy Engine (PE), which grants or denies the request for node access.
To do so, a trustworthiness scheme $\mathcal{M}()$ is established, measuring the mutual trust level of each pair of nodes by a value in [0, 1], i.e.,
\begin{equation}
    \omega_{i \leftrightarrow j} = \omega_{j \leftrightarrow i} = \mathcal{M}(v_i, v_j) \in [0, 1], \forall v_i, v_j \in \boldsymbol{V}^R.
\end{equation}
The higher the $\omega_{i \leftrightarrow j}$ value, the higher the level of trustworthiness that two nodes maintain.
Following zero-trust principles, mutual trustworthiness is updated dynamically, and PE performs the continuous authentication, i.e., the access request will only be granted if the real-time mutual trustworthiness is greater than the user-required threshold.
Finally, PEP executes the trust policies by opening/blocking the communication links according to the PE commands.

\begin{remark}
Since this paper focuses on service provisioning in the zero-trust context, we adopt the general zero-trust policy, i.e., authenticating access according to mutual trustworthiness thresholds. To ensure the adaptability of our proposal, $\mathcal{M}(.,.)$ is regarded as a pluggable module, supporting various well-proven proposals, e.g., reputation \cite{8685209} and tags \cite{tag}. Similarly, more customized/fine-grained authentication mechanisms can be deployed on PE, e.g., blockchain-based voting \cite{10138335}.
\end{remark}

\subsection{Physical Resources Models}
% This paper aims to generate the optimal micro-segmentation catering to the specific task.
% To this end, in this part, we define several critical performance indicators for both the physical and trust layers.
In this part, we define several critical performance indicators for both the physical and trust layers.
\subsubsection{Service Latency}
First, we model the latency required by one micro-segmentation to accomplish the given task.
Denote an $n$-step service as $\mathbf{r}^n = \{r_1, r_2, \dots, r_n\}$. 
$r_j$ ($j \in \{1, 2, \dots, n\}$) represents the $j$-$\mathrm{th}$ step, which calls a certain type of service provider.
%Without loss of generality, we make the following assumptions: 1) $n \leq K$, 2) $s_j \in \{1, 2, \dots, K\}, \forall j \in \{1, 2, \dots, n\}$, and 3) $s_i \textless s_j$, $\forall i, j \in \{1, 2, \dots, n\} \,\&\, i \textless j$.
As aforementioned, $\mathbf{r}^n$ is handled by a $n$-step SFC in the generated micro-segmentation.
The mapping between the zero-trust network and generated micro-segmentation can be expressed as
\begin{equation}
    \mathcal{F}(\boldsymbol{G}^R|\boldsymbol{G}^P) \stackrel{\pi_\theta}{\longmapsto} \boldsymbol{G}^S = (\boldsymbol{C}^n_1, \boldsymbol{C}^n_2, \dots, \boldsymbol{C}^n_Q),
\end{equation}
where $\mathcal{F}(*)$ represents the mechanism to fuse two hierarchical layers. $\boldsymbol{G}^S$ denotes the generated micro-segmentation that contains $Q$ $n$-step SFCs. 
Such a mapping is performed by LEGD, which relies on a learning-based network $\pi$ parameterized by $\theta$.
The LEGD design will be described in Section IV.
When each service request arrives at micro-segmentation $\boldsymbol{G}^S$, it will be randomly assigned to an SFC.

The first component of service latency is performing computation to obtain the required results. 
Denote the computation complexity of $\mathbf{r}^n$ as $\mathbf{c}^n$ = $\{c(r_1), c(r_{2}), \dots, c(r_{n})\}$, where $c(r_{j}) \sim \mathcal{N}\left(\mu^c_{j}, (\sigma^c_{j})^2\right), j \in \{1, 2, \dots, n\}$. 
Hence, the long-term computation latency can be derived as
\begin{equation}
    L_c = \lim_{T\mapsto \infty} \sum\limits_{\mathbf{r}_n \in \mathcal{R}} \sum\limits_{q=1}^{Q}\sum\limits_{j=1}^{n}\frac{\tau\!\left(\pi_{\theta, q}^{-1}(r_j)\right)^{-1}\!\!\!\!c(r_j)}{c\left(\pi_{\theta, q}^{-1}(r_j)\right)}\cdot(TQ)^{-1},
\end{equation}
where $\mathcal{R}$ denotes tasks, arriving following a Poisson distribution.
$\pi_{\theta, q}^{-1}(r_j)$ is the inverse process of Eq. (6), which maps to the device serving $r_j$ in the $q$-$\mathrm{th}$ SFC.
We suppose that the computing power of each device is allocated equally among all SFCs.
$c(*)$ and $\tau(*)$ denote the computing power of device $*$ and the number of SFCs that $*$ belongs to, respectively.

Apart from computation, each service provider in an SFC receives outputs generated by its predecessor and sends its computation results to the successor, causing transmission latency.
Based on \cite{Comm}, the one-hop transmission bandwidth (bit/s) between devices $v_i$ and $v_j \in \boldsymbol{V}^P$ can be defined as
\begin{equation}
    b(e_{i\rightarrow j}) = W\log_2\left(1 + \frac{\varpi(v_i)\,d\left(v_i, v_j\right)^{-\gamma}|h_0|^2}{N_0} \right),
\end{equation}
where $W$ and $\gamma$ are uplink channel bandwidth between devices and path-loss exponent, respectively. 
$h_0$ represents the complex Gaussian channel coefficient following complex normal distribution $\mathcal{CN}(0, 1)$. 
$N_0$ denotes the additive white noise.
Suppose that the bandwidth consumption of intermediate results transferred along the SFC is $\{b(r_{1}), b(r_{2}), \dots, b(r_{n})\}$.
The long-term transmission latency can be derived as
% \begin{equation}
%     L_t = \lim_{T\mapsto \infty} \sum\limits_{\mathbf{r}_n \in \mathcal{R}} \sum\limits_{q=1}^{Q}\sum\limits_{j=1}^{n-1}\Phi_{{j}\leftrightarrow{j+1}}\frac{\tau\!\left(\pi_{\theta, q}^{-1}(e_{j \leftrightarrow j+1})\right)^{-1}b(e_{j \leftrightarrow j+1})}{b\left(\pi_{\theta, q}^{-1}(e_{j \leftrightarrow j+1})\right))}\cdot(TQ)^{-1},
% \end{equation}
\begin{equation}
\begin{aligned}
    L_t &= \lim_{T\mapsto \infty} \sum\limits_{\mathbf{r}_n \in \mathcal{R}} \sum\limits_{q=1}^{Q}\sum\limits_{j=1}^{n-1}\Phi\left(\pi_{\theta, q}^{-1}(r_{j \leftrightarrow j+1})\right) \\
    &\quad\,\,\,\quad\quad \cdot \frac{\tau\!\left(\pi_{\theta, q}^{-1}(r_j)\right)^{-1}b(r_j)}{b\left(\pi_{\theta, q}^{-1}(r_{j \leftrightarrow j+1})\right)} \cdot(TQ)^{-1}.
\end{aligned}
\end{equation}
Similar to Eq. (7), $\pi_{\theta, q}^{-1}(r_{j \leftrightarrow j+1})$ denotes the edge that connects the $j$-$\mathrm{th}$ and $j\!+\!1$-$\mathrm{th}$ service providers in the $q$-$\mathrm{th}$ SFC.
Accordingly, $\Phi\left(\pi_{\theta, q}^{-1}(r_{j \leftrightarrow j+1})\right)$ represents the number of hops existing in this edge.
In this paper, we suppose each pair of service providers directly exchange information without relay, i.e., $\Phi\left(\pi_{\theta, q}^{-1}(r_{j \leftrightarrow j+1})\right)$ = 1 in all the cases.

Combining computation and transmission latency, the total service latency can be expressed as $L_s$ = $L_c + L_t$.

\subsubsection{Service Throughput}
Besides service latency, another critical consideration is the throughput, which refers to the number of service requests that the micro-segmentation can simultaneously process.
Given $Q$ SFCs in the micro-segmentation, the service throughput is $T_s = Q$.

\vspace{-0.2cm}
\subsection{Zero-Trust Model}
\subsubsection{Trustworthiness-aware Service Accomplishment}
%Recall that PEs coordinate the trustworthiness scheme $\mathcal{M}(v_i, v_{j})$ \cite{NIST}, which maintains the real-time mutual trustworthiness of every pair of service providers $\in V^T$.
During the execution of an $n$-step SFC, each service provider $v_{i+1}$ requests $v_{i}$ ($i \in \{1, 2, \dots, n-1\}$) for the intermediate results.
Nonetheless, if mutual trustworthiness $\omega_{i \leftrightarrow i+1}$ is less than the user threshold, PE will deny the access request and terminate the entire SFC execution.
Denote trustworthiness threshold of service $\mathbf{r}^n$ as $\{t(r_1), t(r_2), \dots, t(r_n)\}$, where $t(r_j)\sim N\left(\mu^t_j, (\sigma^t_j)^2\right)$. 
In this case, the probability of successful service execution can be derived as
\begin{subequations}
\begin{flalign}
    P_s & = \sum_{q=1}^{Q}\!\left(\textnormal{Prob}(\boldsymbol{C}_q^n|\boldsymbol{G}^S)\prod_{i=1}^{n-1}\textnormal{Prob}\left(t(r_i) \leq \omega_{i, q} \right)\right) \\
    &= \sum_{q=1}^{Q}\!\left(\prod_{i=1}^{n-1} \frac{Q^{-1}}{\sigma\sqrt{2\pi}}\int_{-\infty}^{\omega_{i, q}}\exp \left(-\frac{(\omega_{i, q} - \mu^t_{i})^2}{2\left(\sigma^t_{i}\right)^2}\right) dt(r_i)\right),
\end{flalign}
\end{subequations}
where $\omega_{i, q}$ represents the mutual trustworthiness between the $i$-$\mathrm{th}$ and $i+1$-$\mathrm{th}$ service provider in the $q$-$\mathrm{th}$ SFC.
By Eq. (10), we can observe that efficient micro-segmentation generation policies should try to configure highly trusted nodes for each step of the SFCs, thus improving the probability of successful service execution.

\subsubsection{Trustworthiness Equilibrium}
Incorporating the mutual trustworthiness of each pair of nodes, the trust layer (see Fig. 1-B) not only indicates the probability of successful service accomplishment but also reflects the network-wide relationship among entities, affecting network stability.
Inspired by equilibrium analysis in signed social networks \cite{SNN}, we present trustworthiness equilibrium to evaluate the stability of the generated micro-segmentation from the trustworthiness perspective.
Trustworthiness equilibrium is defined as
\begin{align}
    E_s = \sum_{q=1}^{Q}\sum_{i=1}^{n}\sum_{j, z} &\frac{|\omega_{i \leftrightarrow j}-\omega_{i \leftrightarrow z}|}{n\,\tbinom{|\boldsymbol{P}_{v_i}|}{2}\,Q}, \forall v_j, v_z \in \boldsymbol{P}_{v_i}, j \textless z, \\
    \boldsymbol{P}_{v_i} &= \{v_k|\exists \,e_{i \leftrightarrow k} \in \boldsymbol{E}^R\},
\end{align}
where $\boldsymbol{P}_{v_i}$ denotes the set of nodes connected to $v_i \in \boldsymbol{G}^R$.
As shown in Fig. \ref{EX}-C, Eq. (11) traverses all nodes in the micro-segmentation.
The difference in mutual trustworthiness values between each node and its two neighbors are accumulated and averaged, thus acquiring $E_s$.
\begin{remark}
We take Fig. \ref{EX}-D as an example of why $E_s$ can reflect the stability.
We can observe that node $v_1$ maintains high trustworthiness with node $v_3$ while is not trusted by node $v_2$.
This discrepancy suggests a potential safety hazard, as some of the node $v_1$'s abnormal behaviors may be hidden by node $v_3$ \cite{SNN}.
Additionally, since trustworthiness values are frequently refreshed in the zero-trust context, high $E_s$ indicates the existence of over-low trustworthiness values, which may cause the nodes to be removed by PE, thus forcing the SFC to be terminated.
\end{remark}
\begin{figure*}[tbp!]
  \centering
  \includegraphics[width=0.92\textwidth]{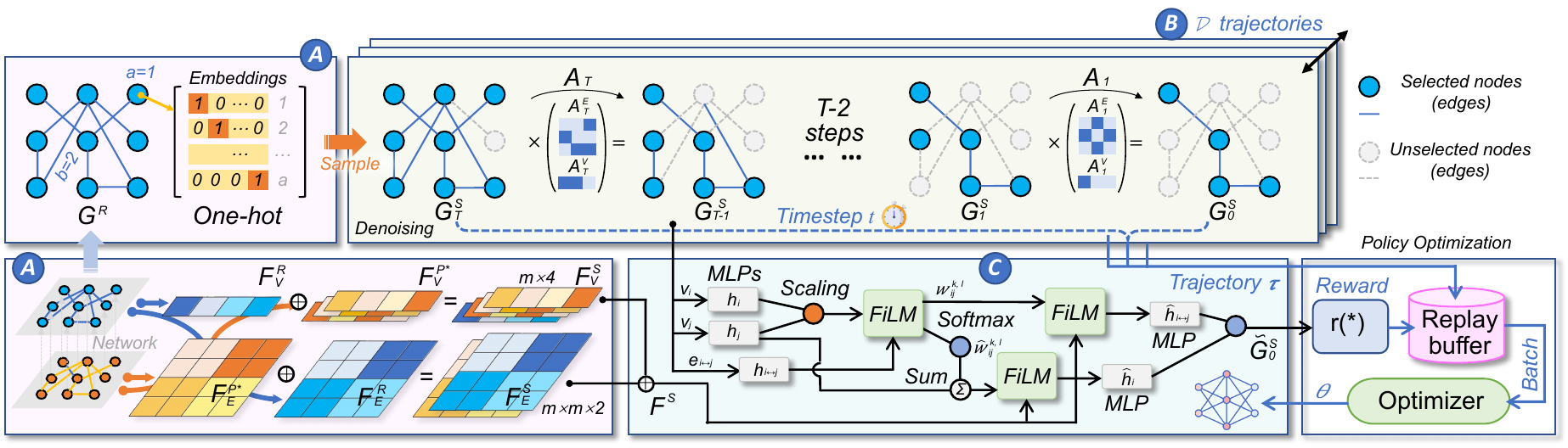} 
  \caption{The illustration of LEGD algorithm. \textbf{A}: The layer fusion and one-hot embeddings of the graph structure. \textbf{B}: The illustration of the denoising process and the trajectory collection module. \textbf{C}: The architecture design of the denoising network. } 
  \label{LEGD}
  \vspace{-0.3cm}
\end{figure*}

\vspace{-0.25cm}
\section{Micro-segmentation Generation via LLM-enhanced Graph Diffusion}
In this section, we formulate the problem of micro-segmentation generation.
We then introduce LEGD, detailing the diffusion paradigm for graph generation, the policy optimization framework, and LLM enhancements.

\subsection{Problem Formulation}
Based on the models in Section III, we formulate the problem for LEGD.
As aforementioned in Eq. (6), we aim to explore an optimal policy for generating micro-segmentations in a controllable way.
Specifically, given the format of service and the zero-trust network modeled by hierarchical graphs $\boldsymbol{G}^P$ and $\boldsymbol{G}^R$, a micro-segmentation $\boldsymbol{G}^S$ should be generated, which accommodates at least one qualified SFC for service provisioning.
Moreover, the generated micro-segmentations should be optimized, i.e., maximizing the overall utility that jointly considers physical- and trust-level performance.
Such a problem can be formulated as
\begin{subequations}
    \begin{flalign}
            \max_{\boldsymbol{G}^S} \;&U_\mathrm{U} \left(\{\boldsymbol{G}^P, \boldsymbol{G}^R, L_s, T_s, P_s, E_s\}\right),\\
            &s.t., ~~~ T_s \geq 1,\\
            &~~~~~~~~L_s \leq L_\mathrm{max},
    \end{flalign}
\end{subequations}
where Eq. (13b) requires that $\boldsymbol{G}^S$ must contain at least one SFC that caters to user services.
Eq. (13c) means the service latency cannot exceed the threshold. 
Otherwise, the service execution will be terminated and regarded as failed.
Jointly considering $L_s$, $T_s$, $P_s$, and $E_s$, the user utility $U_\mathrm{U}$ can be defined as
\begin{equation}
    U_\mathrm{U} = P_s \left[ \alpha_1 \log_{P_s}\left(\frac{L_s}{L_\mathrm{max}}\right) + \alpha_2 T_s + \alpha_3 (1-E_s)\right],
\end{equation}
where $\alpha_1$, $\alpha_2$, and $\alpha_3$ are the weighting factors.
Note that we apply Weber-Fechner Law \cite{FWL} to model the sensitivity of users regarding service latency, which states that the increased latency has a logarithmic relationship with the degradation of user service experience.
Moreover, we adopt $P_s$ as the base of the $\log(\cdot)$ function because the higher the trustworthiness level, the higher the user's tolerance for latency.
Next, we present the LEGD algorithm to solve such a generation problem.

\begin{remark}
    The proposed controllable micro-segmentation generation problem is NP-hard. By regarding the zero-trust network as a partitioning instance and designing the utility function to penalize inter-segmentation edges, the problem can be reduced to the graph partition problem \cite{EDELKAMP2012571}, which is proven to be NP-hard.
\end{remark}

\subsection{Diffusion on Graph for Micro-Segmentation Generation}
We present LEGD, which leverages the diffusion model \cite{diffusion} to achieve strong generation ability and employs an LLM-empowered agent to enhance the capability of topology understanding. Next, we demonstrate the core of LEGD, i.e., diffusion on graph-structured data.

\subsubsection{Forward Diffusion} 
Inspired by non-equilibrium thermodynamics, diffusion consists of two Markov decision processes, namely forward diffusion and denoising.
Denote the optimal micro-segmentation as $\boldsymbol{G}^S_0 = (\boldsymbol{V}^S, \boldsymbol{E}^S)$.
From Eq. (6), we know that $\boldsymbol{G}^S$ is generated by selecting a subset of $\boldsymbol{G}^R$, given the environmental features saved in $\boldsymbol{G}^R$ and $\boldsymbol{G}^P$.
Such a generative problem is discrete, i.e., the nodes and edges should be embedded into categories.
As shown in Fig. 3, we suppose the categories of each node and edge in $\boldsymbol{G}^R$ have $a$ and $b$ possibilities, respectively, and utilize the one-hot embedding to encode such categorical features. 
%Given $\boldsymbol{V}^R \in \mathbb{R}^{m}$ and $\boldsymbol{E}^R \in \mathbb{R}^{m\times m}$, node and edge embeddings can be denoted as $\boldsymbol{V}^S \in \mathbb{R}^{m\times a}$ and $\boldsymbol{E}^S \in \mathbb{R}^{m\times m\times b}$, respectively.

The forward diffusion process gradually disturbs $\boldsymbol{G}^S_0$  into a random graph $\boldsymbol{G}^S_T$ by adding noise incrementally for $T$ steps. Inspired by Digress \cite{digress}, we define noise addition as applying a transition matrix to the current micro-segmentation state, which is suitable for discrete and graph-structured data. Denote the $t$-$\mathrm{th}$ ($t \in \{0, 1, \dots, T\}$) transition matrices for nodes and edges as $\boldsymbol{A}^V_t$ and  $\boldsymbol{A}^E_t$, respectively. The transition probabilities can follow uniform or marginal distributions \cite{digress}. Therefore, for  each node $v_i$ and edge $e_{i \leftrightarrow j} \in \boldsymbol{G}^R$, the category transition at the $t$-$\mathrm{th}$ step is defined as follows:
\begin{subequations}
    \begin{flalign}
    &[\boldsymbol{A}^V_t]_{fg} = q\left(v_{i \mid t+1} = g | v_{i \mid t} = f\right), \\
    &[\boldsymbol{A}^E_t]_{hk} = q\left(e_{i \leftrightarrow j \mid t+1} = k | e_{i \leftrightarrow j \mid t} = h\right),
    \end{flalign}
\end{subequations}
where $f, g \in \{1, \dots, a\}$ and $h, k \in \{1, \dots, b\}$, representing the category of the given node and edge, respectively. 
Hence, the forward diffusion process can be expressed as
\begin{equation}
        q(\boldsymbol{G}^S_T|\boldsymbol{G}^S_0) = \prod_{t=1}^{T} q(\boldsymbol{G}^S_{t}|\boldsymbol{G}^S_{t-1}) = \prod_{t=1}^{T}(\boldsymbol{V}^S\boldsymbol{A}^V_t, \boldsymbol{E}^S\boldsymbol{A}^E_t)
\end{equation}
Denoting $\bar{\boldsymbol{A}}^V = \boldsymbol{A}^V_1\cdot\cdot\cdot\boldsymbol{A}^V_T$ and $\bar{\boldsymbol{A}}^E = \boldsymbol{A}^E_1\cdot\cdot\cdot\boldsymbol{A}^E_T$, Eq. (16) can be rewritten as $(\boldsymbol{V}^S\bar{\boldsymbol{A}}^V, \boldsymbol{E}^S\bar{\boldsymbol{A}}^E)$.

\subsubsection{Denoising}
Denoising can be regarded as the inverse process of forward diffusion, aiming to generate micro-segmentations that can maximize the utility defined in Eq. (14) from a random graph.
Consequently, each denoising step can be expressed as \cite{digress}
\begin{subequations}
    \begin{flalign}
        p_\theta(\boldsymbol{G}^S_{t-1}|&\boldsymbol{G}^S_t) = \!\!\!\!\prod_{1\leq i \leq m}^{}\!\!\!p_\theta(v_i^{t-1}|\boldsymbol{G}^S_t)\!\!\!\!\prod^{}_{1 \leq i, j \leq m}\!\!\!\!p_\theta(e_{i \leftrightarrow j}^{t-1}|\boldsymbol{G}^S_t)\\
 &= \int_{\boldsymbol{G}^S_0}p_\theta(\boldsymbol{G}^S_{t-1}|\boldsymbol{G}^S_t, \boldsymbol{G}^S_0)dp_\theta(\boldsymbol{G}^S_0|\boldsymbol{G}^S_t)\\
 &= \sum_{\widetilde{\boldsymbol{G}}^S_0 \in \mathcal{G}} \underbrace{p_\theta(\boldsymbol{G}^S_{t-1}|\boldsymbol{G}^S_t, \boldsymbol{G}^S_0)}_{\rightarrow q(\boldsymbol{G}^S_{t-1}|\boldsymbol{G}^S_t, \boldsymbol{G}^S_0)}p_\theta(\widetilde{\boldsymbol{G}}^S_0|\boldsymbol{G}^S_t),
    \end{flalign}
\end{subequations}
where $p_\theta(\widetilde{\boldsymbol{G}}^S_0|\boldsymbol{G}^S_t)$ means a prediction on $\boldsymbol{G}^S_0$ given $\boldsymbol{G}^S_t$.
To efficiently represent graph-structured data, we adopt a graph transformer as the denoising network to predict $\boldsymbol{G}^S_0$, which is demonstrated in Section IV-C. 
Since $q(\cdot)$ is pre-defined, $p_\theta(\boldsymbol{G}^S_{t-1}|\boldsymbol{G}^S_t)$ can be calculated.
Hence, we can model the denoising process as a Markov chain, following \cite{Markov}
\begin{equation}
    \begin{split}
        &\boldsymbol{s}_{t} \triangleq \left(\boldsymbol{G}^S_{T-t}, T - t\right), \,\,\boldsymbol{a}_{t} \triangleq \boldsymbol{G}^S_{T-t-1}, \\
        &\pi_{\theta}\left(\boldsymbol{a}_{t} \mid \boldsymbol{s}_{t}\right) \triangleq p_{\theta}\left(\boldsymbol{G}^S_{T-t-1} \mid \boldsymbol{G}^S_{T-t}\right), \\
        &~r\left(\boldsymbol{s}_{t}, \boldsymbol{a}_{t}\right) \triangleq \left\{ \begin{aligned}
            & r\left(\boldsymbol{G}^S_{0}\right), \text{ if } t = T, \\
            & 0, ~~~~~~~~\text{ if } t < T,
        \end{aligned} \right.
    \end{split}
\end{equation}
where $\boldsymbol{s}_t$ and $\boldsymbol{a}_t$ represent the state and action at $t$-$\mathrm{th}$ step, respectively.
$\pi_\theta$ denotes the policy for micro-segmentation and $r(\boldsymbol{s}_t, \boldsymbol{a}_t)$ denotes the reward.
With the Markov decision processes being fixed, we can acquire a series of micro-segmentation generation trajectories, denoted as $\boldsymbol{\tau} = \{\boldsymbol{s}_0, \boldsymbol{a}_0, \boldsymbol{s}_1, \boldsymbol{a}_1, \dots, \boldsymbol{s}_T, \boldsymbol{a}_T\}$.
The accumulative reward of each trajectory is $\sum_{t=0}^{T}r(\boldsymbol{s}_t, \boldsymbol{a}_t) = r(\boldsymbol{G}^S_0)$.
Note that $r(\cdot)$ is defined based on Eq. (14) to measure the performance of the generated micro-segmentation in addressing the optimization problem, which is discussed in Section IV-E.
Then, the expected accumulative reward of the agent can be derived as $\mathcal{J}(\theta) = \mathbb{E}_{p_\theta(\widetilde{\boldsymbol{G^S_0}}|\boldsymbol{G}^S_T)}\left[r(\boldsymbol{G}^S_0)\right]$.

Recall that our goal is to refine denoising network parameters $\theta$, facilitating LEGD to generate the optimal micro-segmentations with the maximum utility, i.e., $\max_{\theta} \mathcal{J}(\theta)$.
Following the policy-based learning principle, we use gradient boosting \cite{boosting} to optimize the policy gradient $\nabla_\theta\mathcal{J}(\theta)$, thereby training LEGD to learn the optimal policy for micro-segmentation generation.
Given $\mathcal{J}(\theta) = \mathbb{E}_{p_\theta(\widetilde{\boldsymbol{G^S_0}}|\boldsymbol{G}^S_T)}\left[r(\boldsymbol{G}^S_0)\right]$, the policy gradient can be derived as 
\begin{subequations}
\begin{flalign}
    \nabla_{\theta} \mathcal{J}_{}(\theta) &=\mathbb{E}_{\pi_\theta}\left[r(\boldsymbol{s}, \boldsymbol{a}) \nabla_\theta \log \pi_\theta(\boldsymbol{a}|\boldsymbol{s})\right] \\
    &=\mathbb{E}_{\boldsymbol{\tau}}\left[r\left(\boldsymbol{G}^S_{0}\right) \sum_{t=1}^{T} \nabla_{\theta} \log p_{\theta}\left(\boldsymbol{G}^S_{t-1} \mid \boldsymbol{G}^S_{t}\right)\right],
\end{flalign}
\end{subequations}
where Eq. (19b) is acquired by substituting $\boldsymbol{s}$, $\boldsymbol{a}$, and $\pi_\theta$ in Eq. (18) to Eq. (19a). Then, we perform Monte Carlo estimation on Eq. (19b), acquiring
\begin{equation}
\begin{split}
     \nabla_\theta \mathcal{J}_{}(\theta) \approx \frac{T}{|\mathcal{D}|\!\cdot\!|\mathcal{T}|} \sum_{\boldsymbol{\tau} \in \mathcal{D}} \sum_{t \in \mathcal{T}} r(\boldsymbol{\tau}.\boldsymbol{s}_0) \nabla_\theta \log p_\theta(\boldsymbol{\tau}.\boldsymbol{s}_{t-1} \mid \boldsymbol{\tau}.\boldsymbol{s}_t),
\end{split}
\end{equation}
where $\mathcal{D}$ and $\mathcal{T}$ represent the sets of sampled trajectories and timesteps, respectively.
$\boldsymbol{\tau}.\boldsymbol{s}_0$ refers to the last generated micro-segmentation in trajectory $\boldsymbol{\tau}$, i.e., $\boldsymbol{G}^S_0$.
In this way, we can efficiently estimate policy gradient and update parameters by interacting with the zero-trust network and generating a batch of trajectories that record the denoising process.
Particularly, to facilitate training converges and alleviate the effect of limited and unreliable Monte Carlo samples, we revise Eq. (20) to \textit{eager policy gradient}, following \cite{GDPO}
\begin{equation}
\begin{split}
     g(\theta) = \frac{T}{|\mathcal{D}|\!\cdot\!|\mathcal{T}|} \sum_{\boldsymbol{\tau} \in \mathcal{D}} \sum_{t \in \mathcal{T}} r(\boldsymbol{\tau}.\boldsymbol{s}_0) \nabla_\theta \log p_\theta(\boldsymbol{\tau}.\boldsymbol{s}_0 \mid \boldsymbol{\tau}.\boldsymbol{s}_t).
\end{split}
\end{equation}
Compared with Eq. (20), eager policy gradient focuses on the prediction of $\boldsymbol{G}^S_0$, thereby reducing the variance of gradient estimates. 
The rationality proof of this strategy can be found in \cite{GDPO}.
Finally, the denoising network parameter $\theta$ can be updated by gradient boosting, i.e., $\theta' = \theta + \eta\cdot g(\theta)$, where $\eta$ represents the learning rate.

% Up till now, we have illustrated the principle of \textit{diffusion on graphs} and derived the training goal.
% Next, we demonstrate the LEGD framework, which accommodates the aforementioned diffusion process and realizes the pipeline of micro-segmentation generation.

\subsection{LEGD Framework}
We present our LEGD framework, consisting of layer fusion, trajectory collection, and the denoising network. Furthermore, we show the data and control flows along these components, which realizes the LEGD training.

\subsubsection{Layer Fusion}
First, we design the input format for LEGD.
As shown in Fig. \ref{LEGD}-A, the inputs of the denoising process, i.e., $\boldsymbol{G}^S_T$, are randomly sampled micro-segmentations.
Recall that micro-segmentation is generated by involving nodes and edges from $\boldsymbol{G}^R$. 
We let $\boldsymbol{G}^S$ share the same size with $\boldsymbol{G}^R$, i.e., $m$ nodes and $m\times m$ edges.
Then, since each edge may be in the \textsl{Selected} or \textsl{Unselected} state while node states are associated with connected edges, we use one-hot encoding with cardinalities $\textit{1}$ and $\textit{2}$ to encode nodes and edges, respectively.
Consequently, we have $v_i \in \mathbb{R}, e_{i \leftrightarrow j} \in \mathbb{R}^{2}$, $\forall v_i, e_{i \leftrightarrow j} \in \boldsymbol{G}^S$, and $\boldsymbol{G}^S_T$ is constructed by randomly initializing these embeddings, as shown in Fig. \ref{LEGD}-A. 

Apart from $\boldsymbol{G}^S_T$, another input for LEGD is the environmental feature $\boldsymbol{F}^S$, which serves as the additional condition of $p_\theta$, i.e., $p_\theta(\boldsymbol{G}^S_{t-1} \!\mid \! \boldsymbol{G}^S_{t}, \boldsymbol{F}^S)$.
We can observe that $\boldsymbol{F}^S$ is the prerequisite to ensure that the generated micro-segmentations can maximize utility in the given network states.
Particularly, both the physical and trust features from the hierarchical graph should be fused and incorporated in $\boldsymbol{F}^S$. 
Hence, we filter $\boldsymbol{G}^P$, deleting all the nodes and corresponding edges that do not exist in $\boldsymbol{G}^R$.
Accordingly, the corresponding node- and edge-level features are also removed from $\boldsymbol{F}^P_V$ and $\boldsymbol{F}^P_E$.
Denoting the filtered $\boldsymbol{G}^P$ as $\boldsymbol{G}^{P*}$, $\boldsymbol{F}^S$ is defined as
\begin{equation}
\boldsymbol{F}^S = \left\{
\begin{aligned}
    &\boldsymbol{F}^S_V \leftarrow \textnormal{Concat}(\boldsymbol{F}^{P*}_V, \boldsymbol{F}^R_V), \\
    &\boldsymbol{F}^S_E \leftarrow \textnormal{Concat}(\boldsymbol{F}^{P*}_E, \boldsymbol{F}^R_E).
\end{aligned}
\right.
\end{equation}
As illustrated by Fig. \ref{LEGD}-A, $ \boldsymbol{F}^{P*}_V \in \mathbb{R}^{m \times 3}$, $\boldsymbol{F}^R_V \in \mathbb{R}^{m \times 1}$, $\boldsymbol{F}^{P*}_E$ and $\boldsymbol{F}^R_E \in \mathbb{R}^{m \times m \times 1}$.
Consequently, we can conclude that $\boldsymbol{F}^S_V \in \mathbb{R}^{m \times 4} $ and $\boldsymbol{F}^S_E \in \mathbb{R}^{m\times m \times 2}$.
Different from $\boldsymbol{G}^S_T$ that initializes the denoising process, $\boldsymbol{F}^S$ is fed into the denoising network for learning $p_\theta(\widetilde{\boldsymbol{G}}^S_0|\boldsymbol{G}^S_t)$, thus facilitating the calculation of $p_\theta(\boldsymbol{G}^S_{t-1}|\boldsymbol{G}^S_t, \boldsymbol{F}^S)$.
The denoising network design is demonstrated below.

\subsubsection{Trajectory Collection}
With $\boldsymbol{G}^S_T$ sampled, the current denoising network performs the denoising process. As illustrated in Fig. \ref{LEGD}-B, once $\boldsymbol{G}^S_0$ is generated, the entire trajectory is saved in a replay buffer, where numerous trajectories are split into multiple training batches. The reward for each trajectory is $r(\boldsymbol{G}^S_0)$. As shown in Eq. (17), to calculate the \textit{eager policy gradient}, we need the prediction of $\boldsymbol{G}^S_0$ given $\boldsymbol{G}^S_t$ and the additional condition $\boldsymbol{F}^S$, i.e., $p_\theta(\boldsymbol{G}^S_{t-1}|\boldsymbol{G}^S_t, \boldsymbol{F}^S)$. To achieve this, we randomly sample timesteps from ${1, 2, \dots, T}$ for predictions, reducing computation overhead and avoiding model overfitting \cite{diffusion}. The LEGD algorithm continuously interacts with the zero-trust network, denoising different $\boldsymbol{G}^S_T$ and $\boldsymbol{F}^S$, thus enriching the replay buffer. A batch of trajectories is fetched at each epoch, and the \textit{eager policy gradient} is calculated to update model parameters.

\subsubsection{Denoising Network}
The denoising network is the learnable component of LEGD, which predicts the clean micro-segmentation $\boldsymbol{G}^S_0$ based on input $\boldsymbol{G}^S_t$ and condition $\boldsymbol{F}^S$.
As shown in Eq. (17c), two requirements should be satisfied since the prediction of $\boldsymbol{G}^S_0$ directly determines the performance of micro-segmentation generation.
First, the denoising network should effectively represent complicated topologies and graphical features, learning the graph generation policy.
Moreover, it should support addition conditions (i.e., $\boldsymbol{F}^S$), thus enabling the controllable generation for maximizing utilities.
To this end, we build the denoising network for LEGD based on graph transformer architecture \cite{GDPO, GTN}.

As shown in Fig. \ref{LEGD}-C, nodes $\boldsymbol{V}^S_t$ and edges $\boldsymbol{E}^S_t$ first undergo the corresponding Multi-layer Perception (MLP) modules \cite{digress} to acquire the embeddings.
Afterward, a graph transfer layer is deployed, incorporating the graph attention mechanism.
Through graph attention scores, graphical features, especially the message transfers among nodes along edges, can be reflected.
Similar to the conventional attention \cite{transformer}, the score of the $\ell$-$\mathrm{th}$ attention layer is defined as
\begin{subequations}
    \begin{flalign}
        \hat{\boldsymbol{w}}_{i j}^{k, \ell}= \textnormal{FiLM}&\left(\left(\frac{\boldsymbol{Q}^{k, \ell} \boldsymbol{h}_{i}^{\ell} \cdot \boldsymbol{K}^{k, \ell} \boldsymbol{h}_{j}^{\ell}}{\sqrt{d_{k}}}\right)\!, \boldsymbol{E}^{k, \ell} \boldsymbol{h}_{i\leftrightarrow j}^{\ell}\right),\\
        &~~~\boldsymbol{w}_{i j}^{k, \ell} = \textnormal{softmax}(\boldsymbol{\hat{w}}_{i j}^{k, \ell}),
    \end{flalign}
\end{subequations}
where $k \in \{1, 2, \dots, H\}$ denotes the index of attention head.
$\boldsymbol{h}_i$ and $\boldsymbol{h}_{i \leftrightarrow j}$ represent the embeddings of $v_i$ and $e_{i \leftrightarrow j} \in \boldsymbol{G}^S$, respectively.
Eq. (23a) shows that for each node, its attention to all neighbors and the corresponding edges will be incorporated, thus informing the denoising network with graph-perspective knowledge.
Note that FiLM refers to the \textit{Feature-wise Linear Modulation} layer \cite{FiLM}, which can modify the output of the transformer layer by applying a scaling and shifting operation that is conditioned on $\boldsymbol{F}^S$.
Hence, the processed node and edge features can be defined as
\begin{subequations}
    \begin{flalign}
        ~~\boldsymbol{\hat{h}}_{i} & = \textnormal{FiLM}\left(\|_{k=1}^{H}\left(\sum_{j} \boldsymbol{w}_{i j}^{k, \ell} \boldsymbol{V}^{k, \ell} \boldsymbol{h}_{j}^{\ell}\right), \boldsymbol{F}^S\right), \\
        &~~~\boldsymbol{\hat{h}}_{i \leftrightarrow j} =\textnormal{FiLM}\left( \|_{k=1}^{H}\left(\boldsymbol{\hat{w}}_{i j}^{k, \ell}\right), \boldsymbol{F}^S\right),
    \end{flalign}
\end{subequations}
where $||$ represents the Concat operation that combines the outputs of $H$ attention heads.
Afterward, the processed node and edge features go through residual connection and layer normalization \cite{transformer}, which are utilized to combat the vanishing gradient problem and stabilize the training process, respectively.
Finally, two MLPs are deployed to decode node and edge, respectively.
In this way, the embeddings can be transferred to node and edge matrices that correspond to the predicted micro-segmentation $\boldsymbol{G}^S_0$.
%\begin{figure}[tbp!]
%  \centering
%  \includegraphics[width=0.47\textwidth]{LLM-2.pdf} 
%  \caption{The LLM enhancement in LEGD.} 
%  \label{ENHANCE}
%\end{figure}

\subsection{Reward Engineering}
To provide efficient feedback for every generated $\boldsymbol{G}^S_0$ and guide the LEGD training, the reward function $r(\cdot)$ should be crafted, so-called reward engineering.
Considering both the objective function and constraints, we present the reward function that considers both the direct reward and the penalty terms.
The former reflects the unconstrained user utility, and the latter ensures compliance with the constraints. 
To harmonize the relationship between these two aspects, we define a penalty-constrained reward expression, i.e.,
\begin{equation}
\begin{split}
        r(\boldsymbol{G}^S_0) = {U_\mathrm{U}}\left(\{\boldsymbol{G}^P, \boldsymbol{G}^R, L_s, T_s, P_s, E_s\}\right) \\\times \underbrace{(\Omega_{\rm T} \times \Omega_{\rm L})}_{\textnormal{penalty terms}},
\end{split}
\end{equation}
where $\Omega_{\rm T} = \mathcal{I}(T_s, 1)$ and $\Omega_{\rm L} = \mathcal{I}(L_\mathrm{max}, L_s)$.
Note that $\mathcal{I}(*, \circ)$ denotes the indicator function, which outputs \textit{1} if $* \geq \circ$ and \textit{0} otherwise.
Substituting Eq. (25) into Eq. (21) and establishing the framework illustrated in Section IV-C, the basic LEGD algorithm is completely demonstrated and can address the controllable micro-segmentation generation problem defined by Eq. (13).
\begin{algorithm}[tpb]
    {\footnotesize \caption{\textcolor{black}{The proposed LEGD approach}} \label{alg:3}
    $\mathbf{Input}$: LLM-empowered agent $\mathcal{A}_\mathrm{LEGD}$, initial denoising network $\pi_\theta$, network structure and features $\boldsymbol{G}^R$ and $\boldsymbol{G}^{P*}$, temperature value $\varphi$, \# of diffusion steps $T$, \# of timestep samples $|\mathcal{T}|$, \# of trajectory sample $\mathcal{D}$, learning rate $\eta$, \# of training steps $N_t$;\\
    \vspace{0.5em}
    \textit{Procedure 1: LLM-based Enhancement};\\
    \quad Inform LLM with $\boldsymbol{G}^R$ and $\boldsymbol{G}^{P*}$\\
    \quad Acquire filters $\boldsymbol{f}$ = \{$f_1, f_2, \dots, f_n$\}\\
    \quad $\boldsymbol{G}^{S*}_T = \Psi_{\varphi}\left(\left(\prod_{i=1}^{n}f_i\right)\boldsymbol{G}^S_T\right)$\\
    \vspace{0.5em}
    \textit{Procedure 2: LEGD Training};\\
    \quad  \For{$i = 1, 2, \dots, N_t$}{    
    \quad    \For{$d = 1, 2, \dots, |\mathcal{D}|$}{
    \quad     Sample $\boldsymbol{G}^{S*}_T$ based on \textit{Procedure 1}\\
    \quad        \For{$t = 1, 2, \dots, T$}{
    \quad                Perform denoising based on Eq. (17)
    } 
    \quad Acquire trajectory $\boldsymbol{\tau}_d$ \\
    \quad Sample timestep $T_d$ $\sim$ $\mathrm{Uniform}([1, T])$\\
    \quad Call $\pi_\theta$ to predict $\boldsymbol{G}^S_0$\\
    \quad Calculate reward $r(\widetilde{\boldsymbol{G}}^S_0)$ based on Eq. (25)\\
    }
    }
    \quad calculate reward mean $\bar{r} \leftarrow \frac{1}{|\mathcal{D}|} \sum_{d=1}^{|\mathcal{D}|} r_d$\\
    \quad calculated reward variance $\text{std}[r] \leftarrow \sqrt{\frac{1}{|\mathcal{D}|-1}\sum_{d=1}^{|\mathcal{D}|} (r_d - \bar{r})^2}$\\
    \quad calculate eager policy gradient $g(\theta)$ based on $\bar{r}$, $\text{std}[r]$, and Eq. (21)\\
    \quad Update $\pi_\theta$ parameters\\
    \vspace{0.5em}
    \textit{Procedure 3: LEGD Inference};\\
    \quad Sample $\boldsymbol{G}^{S*}_T$ based on \textit{Procedure 1}\\
    \quad \For{$t = 1, 2, \dots, T$}{
    \quad Perform denoising based on Eq. (17)\\
    }
    $\mathbf{Output}$: Generated micro-segmentation $\boldsymbol{G}^S_0$
}\end{algorithm}

\subsection{LLM Enhancement}

Similar to GNNs, LEGD may take a long time to converge as the denoising network struggles to capture the graphical features and generate optimal micro-segmentations in varying environments. With the increasing network scale and complexity in the NGN era \cite{10437338}, we aim to improve LEGD's efficiency. Inspired by Reinforcement Learning with Human Feedback (RLHF) \cite{RLHF}, we exploit human perception and expertise. While humans cannot directly generate optimal micro-segmentations due to numerous possibilities and intensive calculations, they excel at recognizing patterns and identifying outliers \cite{Pattern}. This allows us to remove certain graph parts and simplify the generation problem. For example, in Fig. \ref{TR}-A, humans can easily filter out text-to-3D service providers since they are not required by users. In contrast, neural networks require numerous samples and efficient data representation to learn such patterns. Conventional RLHF relies heavily on real-time human feedback, which is labor-intensive and unstable. Fortunately, LLMs, trained on massive datasets, possess human-like expertise and multimodal understanding. To this end, we leverage an LLM-empowered agent to simulate human perception and reduce LEGD's action space. Specifically, the agent is informed with the following information
\begin{equation}
    \mathcal{A}_\mathrm{LEGD} = \textnormal{LLM}(\boldsymbol{f}\,|\,\boldsymbol{G}^S_T, \boldsymbol{G}^{P*}, \boldsymbol{G}^R, U_\mathrm{U}),
\end{equation}
where $\boldsymbol{G}^S_T$ informs LLM with the original state space structure. 
Graphs $\boldsymbol{G}^{P*}$ and $\boldsymbol{G}^R$ provide the environmental context, i.e., the features of each node and edge.
$U_\mathrm{U}$ indicates the requirements for the generated micro-segmentations.
Then, $\mathcal{A}_\mathrm{LEGD}$ can analyze the aforementioned information and generate a series of heuristic features $\boldsymbol{f} = \{f_1, f_2, \dots, f_n\}$.
Each filter corresponds to one rule that removes certain nodes' candidacy for constructing micro-segmentation.
Thus, the state space of LEGD can be effectively shrunk by applying these filters, i.e.,
\begin{equation}
    \boldsymbol{G}^{S*}_T = \Psi_{\varphi\sim[0, 1]}\left(\left(\prod_{i=1}^{n}f_i\right)\boldsymbol{G}^S_T\right),
\end{equation}
% where $\Psi$ denotes the temperature mechanism, which controls the strictness of compliance with $\mathcal{A}_\mathrm{LEGD}$'s rules.
% $\varphi$ = 1 means all the nodes filtered by $\mathcal{A}_\mathrm{LEGD}$ are removed, which can minimize the state space.
% In contrast, a lower $\varphi$ value provides better exploration ability to LEGD since the policies generated by $\mathcal{A}_\mathrm{LEGD}$ may not be perfect.
% In the aforementioned LEGD framework, $\mathcal{A}_\mathrm{LEGD}$ is deployed online within the trajectory collection module.
% Finally, to ensure the adaptability of LEGD, $\mathcal{A}_\mathrm{LEGD}$ is pluggable.
% Any LLM that owns graph understanding capability can be applied, such as ChatGPT-4 \cite{chatgpt} or LLaVA \cite{llava}.
% The entire process of LEGD is shown in \textbf{Algorithm 1}.
where temperature mechanism $\Psi$ controls the strictness of compliance with $\mathcal{A}_\mathrm{LEGD}$'s rules. When $\varphi = 1$, all nodes filtered by $\mathcal{A}_\mathrm{LEGD}$ are removed, minimizing the state space. 
Lower $\varphi$ values allow better exploration, as the filters adopted by $\mathcal{A}_\mathrm{LEGD}$ may not be perfect. 
To ensure adaptability, $\mathcal{A}_\mathrm{LEGD}$ is designed as a pluggable module with the trajectory collector.
Any LLM with graph understanding capability, such as ChatGPT-4 \cite{chatgpt} or LLaVA \cite{llava}, can be applied. 
Moreover, $\mathcal{A}_\mathrm{LEGD}$ works in an online manner during the LEGD training, thereby dynamically configuring and activating filters. 
The entire workflow of LEGD is detailed in \textbf{Algorithm 1}.

\section{Adaptive Micro-segmentation Maintenance}
LEGD realizes the optimal micro-segmentation generations for the specific network states.
Nonetheless, in zero-trust NGN, two situations may occur.
\begin{itemize}
     \item \textbf{Trustworthiness Update}: Zero-trust paradigms involve continuous trustworthiness updates \cite{10052642}. As presented in Section III, an increase in trustworthiness benefits the service experience. In contrast, if one entity loses the trust of neighbors due to suspicious behaviors, the execution success rate of its SFC will be sharply reduced. Moreover, if its mutual trustworthiness drops below the threshold, it might be removed by PE, causing the failure of the entire micro-segmentation.
     \item \textbf{Service Upgrade}: The services in the NGN era are rapidly evolving. Taking AIGC as an example, the emergence of ChatGPT-4, which supports conversational graph generations, greatly reduces the demand for text-to-image services, such as Stable Diffusion. Hence, a topology of micro-segmentation should evolve according to the changing service requirements.
\end{itemize}
Both two situations require dynamic micro-segmentation maintenance, i.e., keeping the overall micro-segmentation topology while involving/removing a part of nodes/edges to adapt to the new environment/services.
To this end, we further present LEGD-AM, which performs task-oriented fine-tuning atop LEGD to fit varying zero-trust networks.
%LEGD optimizes micro-segmentation generation for specific network states. However, two frequent situations in zero-trust NGN require attention: {\textbf{topology updates}} and {\textbf{service updates}}. First, continuous trustworthiness updates may affect service execution success and potentially cause micro-segmentation failure if trustworthiness drops below a threshold \cite{10052642}. Additionally, rapidly evolving services, such as the shift in demand from text-to-image services to conversational graph generations with ChatGPT-4, necessitate changes in micro-segmentation topology \cite{SD}. To address these challenges, we present LEGD-AM, which fine-tunes LEGD to adapt dynamically to varying zero-trust network environments.
\begin{figure*}[tbp!]
  \centering
  \includegraphics[width=\textwidth]{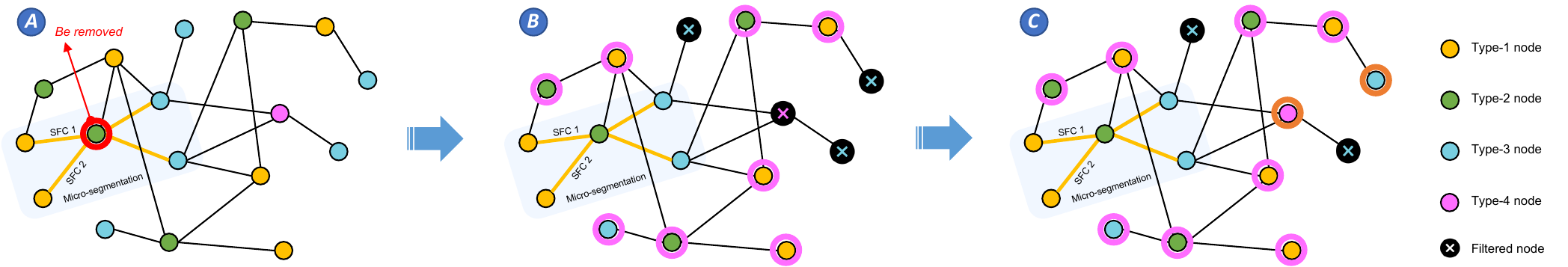} 
  \caption{The illustration of an interest zone. \textbf{A}: The zero-trust network and the original micro-segmentation. Note that a type-\textit{2} node is removed due to over-low trustworthiness. \textbf{B}: 1-degree interest zone, which is highlighted by pink circles, including all remaining type-\textit{2} nodes and their 1-degree neighbors. \textbf{C}: 2-degree interest zone, adding the 2-degree neighbors, which are highlighted in orange. The nodes marked by black crosses will be marked and filtered by the adaptive mask.} 
  \label{INT}
\end{figure*}

\subsection{Fine-tuning of LEGD}
First, we utilize the well-trained LEGD model (parameterized by $\theta$) as the checkpoint.
Suppose that after the LEGD training converges, the model can efficiently generate micro-segmentation for \{type-\textit{1}, type-\textit{2}, type-\textit{3}\} services.
During the operation, one type-\textit{2} device is removed, or the micro-segmentation needs to upgrade the service type to \{type-\textit{1}, type-\textit{2}, type-\textit{3}, type-\textit{4}\}, we should involve/remove certain nodes (and the corresponding edges) while retaining the overall topology.
Hence, we leverage the fine-tuning technique \cite{10054400} to align the pre-trained model with new objectives.
The fine-tuning process of LEGD can be expressed as
\begin{equation}
    \theta^* = \theta + \eta_f g_{\theta}\left(\theta; r_f(\boldsymbol{G^{S*}_0} \mid \boldsymbol{G}^S_0)\right),
\end{equation}
where $\eta_f$ represents the fine-tuning learning rate. 
$\boldsymbol{G}^{S*}_0$ means the refined micro-segmentation based on the new objective, which is expressed by the fine-tuning reward function $r_f(*)$.
$g_{\theta}\left(\theta; r_f(\boldsymbol{G^{S*}_0} \mid \boldsymbol{G}^S_0)\right)$ denotes the eager policy gradient for fine-tuning.
Compared with training a model from scratch, fine-tuning requires much fewer epochs to converge since the pre-trained knowledge can be utilized.

\subsection{LEGD-AM Design}
\subsubsection{Adaptive Mask}
To perform fine-tuning, first, we present adaptive masks to constrain the structure of $\boldsymbol{G}^{S*}_0$, thereby aligning it to the fine-tuning objective.
Masks act similarly to the filters mentioned in Section IV, which disables a part of nodes' candidacy for participating in $\boldsymbol{G}^{S*}_0$ during the entire denoising process.
Such an operation can improve training efficiency by allowing the denoising network $\pi$ to selectively focus on the most relevant graph parts.
Moreover, it can minimize the number of nodes and edges that can be adjusted, thus reducing micro-segmentation costs.
This is because the involvement/removal of each node requires PE to redeploy the network virtualization, re-evaluate the trustworthiness, etc. \cite{Micro2}.
To develop adaptive masks for LEGD-AM, we first define the concept of an $\epsilon$-\textit{Interest Zone}.
\begin{definition}
($\epsilon$-Interest Zone). For given graph $\boldsymbol{G}^R$, an $\epsilon$-interest zone is defined as the subset of $\boldsymbol{G}^R$, which includes all nodes (and the corresponding edges) that provide the required type of services and their $\epsilon$-degree neighbors.
\end{definition}

Fig. \ref{INT} demonstrates the meaning of $\epsilon$-interest zone. 
Suppose that one type-\textit{2} node is removed due to over-low trustworthiness.
In this case, to fix the micro-segmentation, the required service type is \textit{2}.
The corresponding \textit{1}- and \textit{2}-interest zones, denoted by $\mathbb{Z}^1$ and $\mathbb{Z}^2$, respectively, are illustrated in Fig. \ref{INT}-B and Fig. \ref{INT}-C.
Accordingly, masks filter the graph parts outside the interest zone and can be designed as
\begin{equation}
    \boldsymbol{M} \in \mathbb{R}^{m \times m}, \,\, [\boldsymbol{M}]_{ij} = [\boldsymbol{M}]_{ji} =
    \begin{cases}
        1, & \text{if}\, f^R_{v_i} \notin \mathbb{Z}^{\epsilon} \\
        0, & \text{otherwise}
    \end{cases},
\end{equation}
where $i$, $j \in \{1, 2, \dots, m\}$.
Similar to Eq (27), the adaptive masks are applied to $\boldsymbol{G}^S_T$. 
Moreover, a temperature mechanism like $\Psi$ can also be applied to balance training complexity and exploration capability.

\subsubsection{Adaptive Reward Engineering}
The reward $r_f(*)$ should be crafted to guide fine-tuning, effectively indicating the desirability of each action in refining the micro-segmentation.
Considering that fine-tuning aims to balance the utility of the update micro-segmentations and the re-configuration costs, we define the following two-term reward for LEGD-AM:
\begin{equation}
    r_f(\boldsymbol{G}^{S*}_0\mid \boldsymbol{G}^S_0) = U_\mathrm{U} - \underbrace{\alpha_4 \Delta(\boldsymbol{G}^S_0, \boldsymbol{G}^{S*}_0)}_{\text{re-configuration costs}},
\end{equation}
where $U_\mathrm{U}$ is defined in Eq. (14) and represents the utility of the refined micro-segmentation $\boldsymbol{G}^{S*}_0$.
$\alpha_4$ is a weighting factor.
$\Delta(\boldsymbol{G}^S_0, \boldsymbol{G}^{S*}_0)$ measures the topological difference between the original micro-segmentation $\boldsymbol{G}^S_0$ and the refined one $\boldsymbol{G}^{S*}_0$, which can be calculated using well-established metrics such as \textit{graph edit distance} \cite{lv2023graph}.
From Eqs. (29) and (30), we can observe that LEGD-AM ensures the adaptability of our service provisioning framework in zero-trust NGN since both the mask and reward function can be flexibly designed according to the latest environment/requirements.

\section{Numerical Results}
In this section, we implement the proposed zero-trust service provisioning framework and the micro-segmentation generation and update algorithms.
Then, we conduct extensive experiments that aim to answer: 1) whether the LEGD algorithm can generate the micro-segmentations that maximize utility $U_\mathrm{U}$ and 2) whether the LEGD-AM algorithm can efficiently maintain micro-segmentation topologies according to the varying network environments.

\textbf{Experimental Settings.} The experiments are conducted on a server with an NVIDIA RTX A5000 GPU with 24 GB of memory and an AMD Ryzen Threadripper PRO 3975WX 32-Core CPU with 263 GB of RAM in Ubuntu 20.04 LTS. We utilize this server to simulate a nine-node zero-trust AIGC network. We suppose that the network contains four types of service providers, namely text generation, text-to-image synthesis, image-to-video creation, and interactive video enhancement. These types are indexed by $\textit{1}$, $\textit{2}$, $\textit{3}$, and $\textit{4}$, respectively. Users request text-controlled video generation services, i.e., \{type-\textit{1}, type-\textit{2}, type-\textit{3}\}. The zero-trust architecture is built following the NIST standard \cite{NIST}. Additionally, we adopt reputation \cite{8685209} and ChatGPT-4 \cite{chatgpt} as the trustworthiness scheme and LLM-empowered agent, respectively. 
% Finally, The proposed service provisioning framework and algorithms are implemented in Python with PyTorch 2.0.1. 
\begin{figure}[tbp!]
  \centering
  \includegraphics[width=0.5\textwidth]{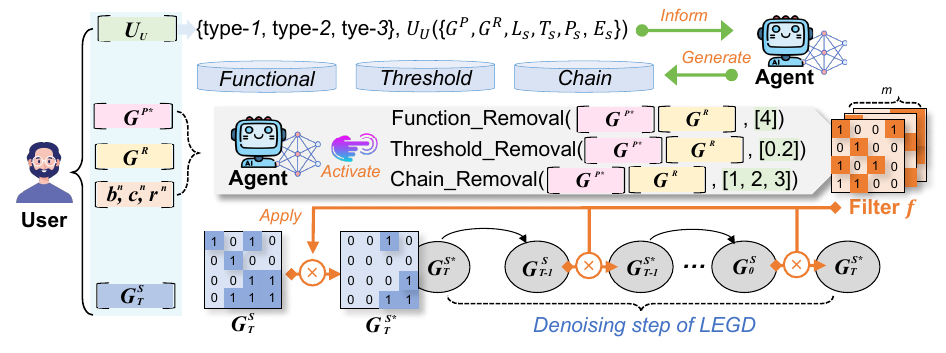} 
  \caption{The generation and activation of heuristic filters by the LLM-empowered agent.} 
  \label{LLM}
\end{figure}

\begin{figure}[tbp!]
  \centering
  \includegraphics[width=0.49\textwidth]{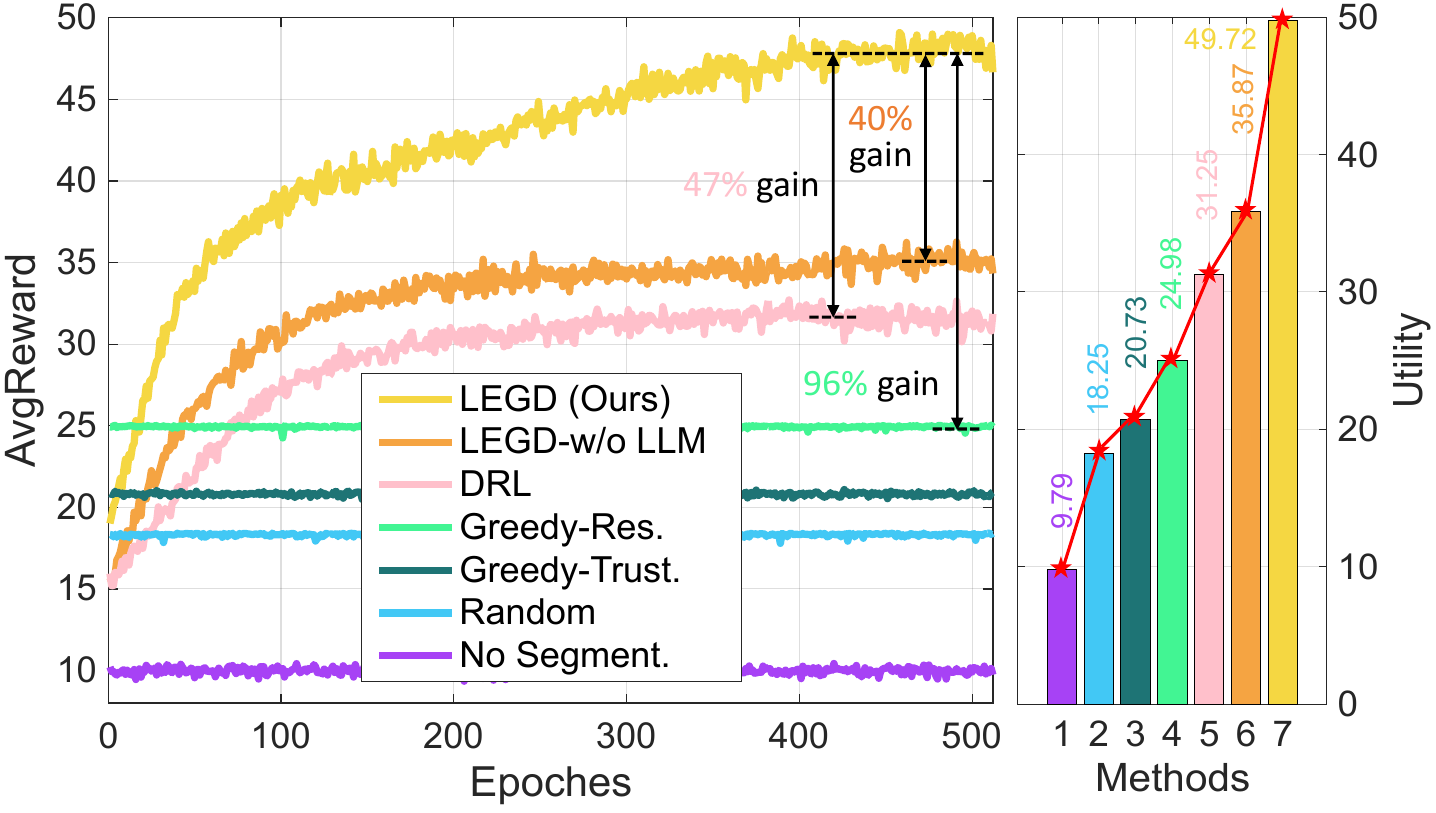} 
  \caption{The AvgReward of LEGD and other baselines over training epochs.} 
  \label{TEST1}
\end{figure}
\begin{figure*}
\centering
\begin{minipage}{\textwidth}
	\centering
	\begin{minipage}{0.31\textwidth}
	\includegraphics[width=\textwidth]{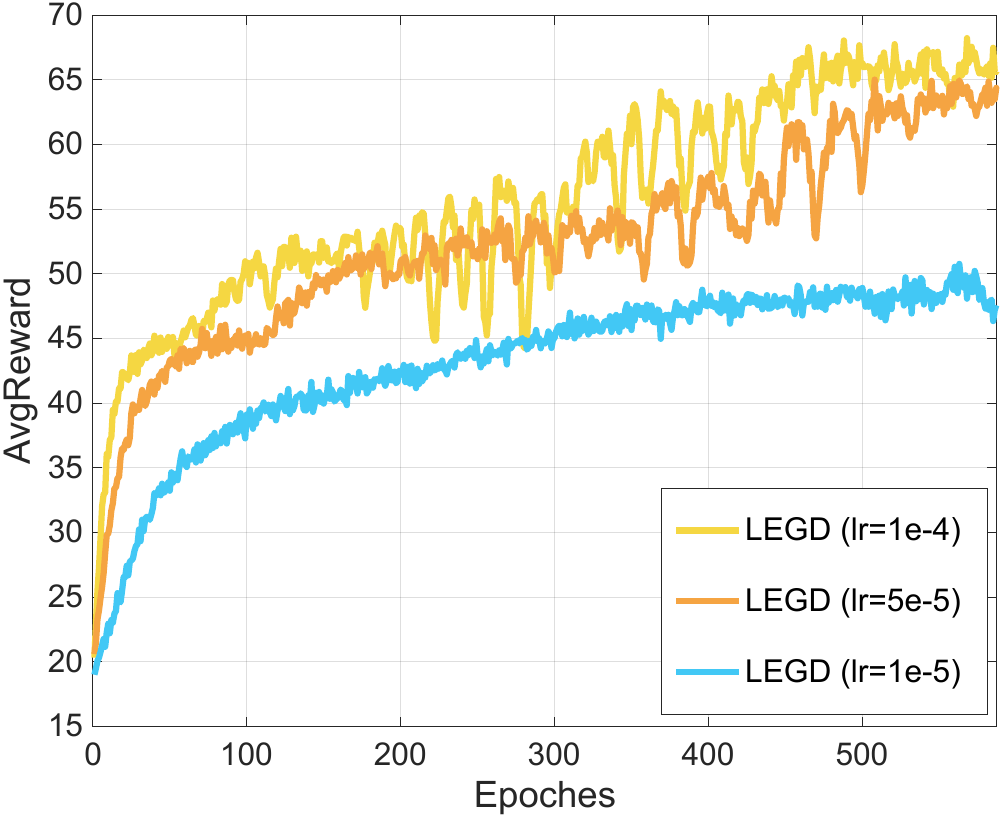}
	\subcaption*{(a) The impact of learning rate.}
	\end{minipage}
	\begin{minipage}{0.31\textwidth}
	\includegraphics[width=\textwidth]{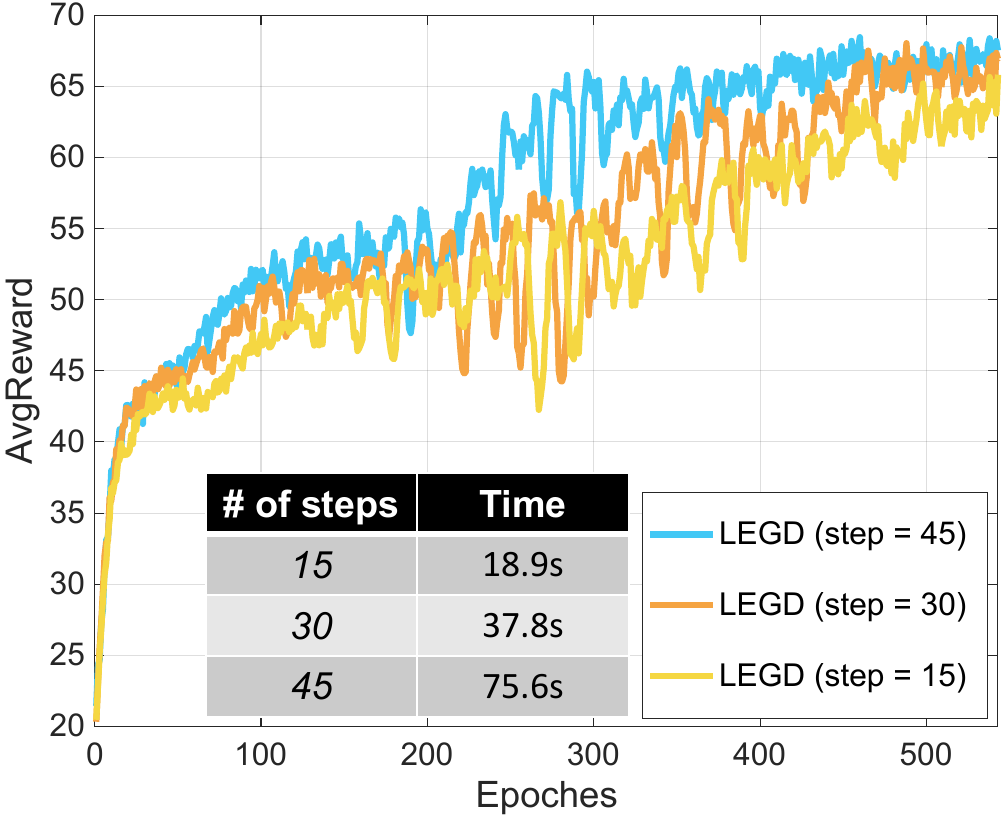}
        \subcaption*{(b) The impact of diffusion step.}
	\end{minipage}
	\begin{minipage}{0.31\textwidth}
	\includegraphics[width=\textwidth]{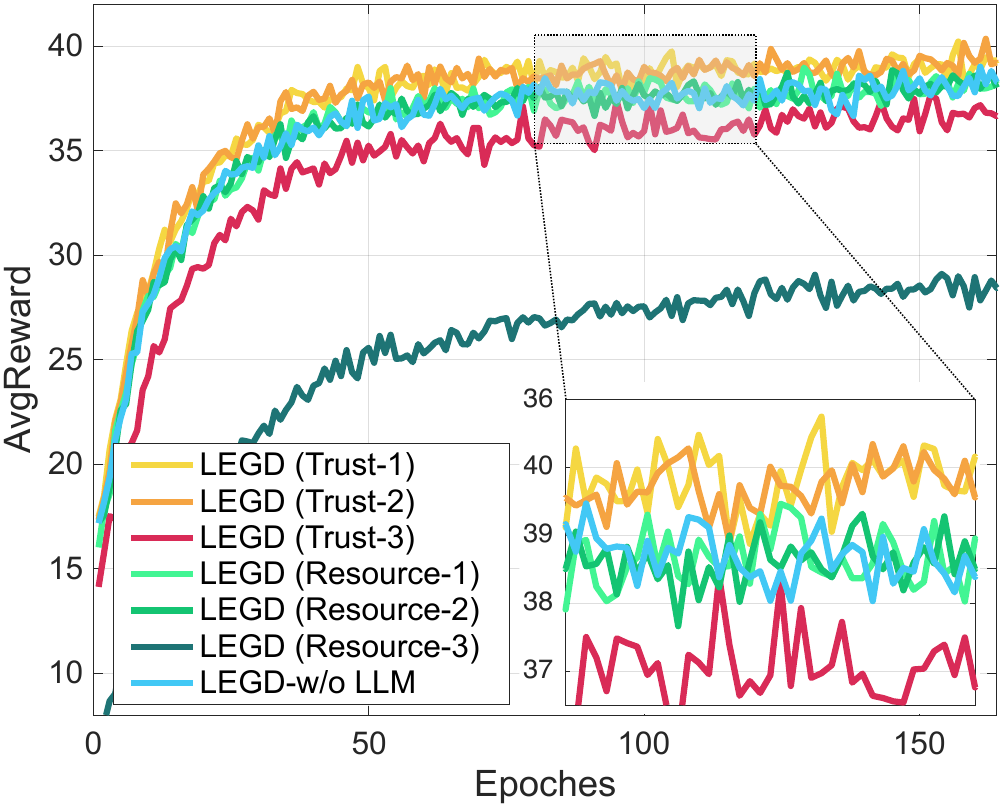}
	\subcaption*{(c) The impact of filter activation.}
	\end{minipage}
        \caption{The impact of LEGD configurations on AvgReward. Note that for subfigure (c), resource-1, -2, and -3 refer to removing edges with mutual trustworthiness lower than 0.1, 0.3, and 0.5, respectively. In addition, resource-1, -2, and -3 refer to removing nodes with [computing, transmission] power less than [50GFLOPS, 20W], [70GFLOPS, 30W], and [90GFLOPS, 40W], respectively.}
	\label{TEST2}
\end{minipage}
 \vspace{-0.2cm}
\end{figure*}

\subsection{Filter Generation and Dynamic Activation}
%\subsubsection{Agent Training}
%First, we train the LLM-empowered agent to generate heuristic filters for LEGD. As shown in Fig. \ref{LLM}, we feed it with the detailed system model and problem formulation. Contributed to the outstanding reasoning capability, LLM can associate the environment with objectives, thus outputting various heuristic filter designs.

%\subsubsection{Heuristic Filters Generation}
%According to our experimental settings, the LLM presents three heuristic filters: 1) \textbf{Remove-4}: Remove all the edges connected to type-$\textit{4}$ nodes; 2) \textbf{Trust-oriented}: Remove all the edges whose associated mutual trustworthiness is less than the pre-defined threshold, and 3) \textbf{Resource-oriented}: Remove all the edges whose connected nodes have resources (including computing and transmission power) less than the threshold.

%Note that more application-specific filters can be generated by feeding the LLM with more background information or fine-tuning its answer.
%Since our experiments mainly intend to prove the effectiveness of introducing heuristic expertise knowledge in graph generation, we use these three basic filters.
%In the next part, we will compare the performance between pure graph diffusion and LEGD. }
The LLM-empowered agent improves LEGD by generating and dynamically activating multiple filters. Based on the graphic network modeling (in Section III) and the utility function for micro-segmentation generation (in Section IV), the heuristic filters provided by the agent include:
\begin{itemize}
\item \textbf{Functional Filter}: Targets and removes nodes that do not meet current service requirements, streamlining the network structure to better fit operational needs.
\item \textbf{Threshold Filter}: Evaluate nodes and edges based on real-time assessments of resource capacity and trustworthiness, selectively removing those with low resources or trustworthiness during the denoising process.
\item \textbf{Chain Filter}: Analyzes and excludes unnecessary service chains, optimizing the path of service function chains to enhance overall network efficiency.
\end{itemize}

As shown in Fig. \ref{LLM}, during LEGD training, the agent keeps monitoring the dynamic user requirements, including service type, computation complexity $\mathbf{c}^n$, bandwidth consumption $\mathbf{b}^n$, and trustworthiness thresholds $\mathbf{r}^n$, and manages the filter activation accordingly.
For example, agent's decision \{``Function'': [4], ``Trust'': [0.2], ``Chain'':[1, 2, 3]\} means activating the following mechanisms to form a combined filter that: 1) Remove any connection with type-\textit{4} nodes; 2) Remove the edges where the trustworthiness is lower than 0.2; and 3) Remove connections outside service chains \{type-\textit{1}, type-\textit{2}, type-\textit{3}\}.
This setup enables LEGD to combine the neural network's learning capability and heuristic expertise knowledge, thus further improving efficiency.
Note that more application-specific filters can be generated to accommodate more specific service requests based on user's preferences.
Since our experiments mainly intend to demonstrate the effectiveness of LLM enhancement, we adopt three basic filters mentioned above.
Next, we compare the performance of LEGD with and without LLM enhancement. 

\subsection{Efficiency of LEGD}
\subsubsection{Baseline Settings}
We adopt the following baselines: \textbf{No-Segmentation}, where the entire zero-trust network forms a single micro-segmentation and service requests are randomly assigned to one SFC; \textbf{Random}, where the zero-trust network is randomly divided into multiple micro-segmentations and service requests are randomly assigned to one SFC within a micro-segmentation; \textbf{Greedy-Resource}, which uses Metis \cite{Metis}, a widely adopted algorithm for segmenting graphs with minimal costs, for micro-segmentation generation, setting mutual trustworthiness as the cost to reflect the greedy strategy for maximizing mutual trustworthiness of generated micro-segmentations; \textbf{Greedy-Trust}, which also uses Metis but sets the total volume of resources (including both the computing and transmission power) as the costs, reflecting the greedy strategy for maximizing the total resources of generated micro-segmentations; and \textbf{DRL}, which follows the conventional policy-based DRL paradigm \cite{10309220} rather than generative architectures. Note that to ensure fairness, the neural network adopted by DRL for graph representation and understanding is the same as that of LEGD.

\begin{figure*}[tbp!]
  \centering
  \includegraphics[width=0.92\textwidth]{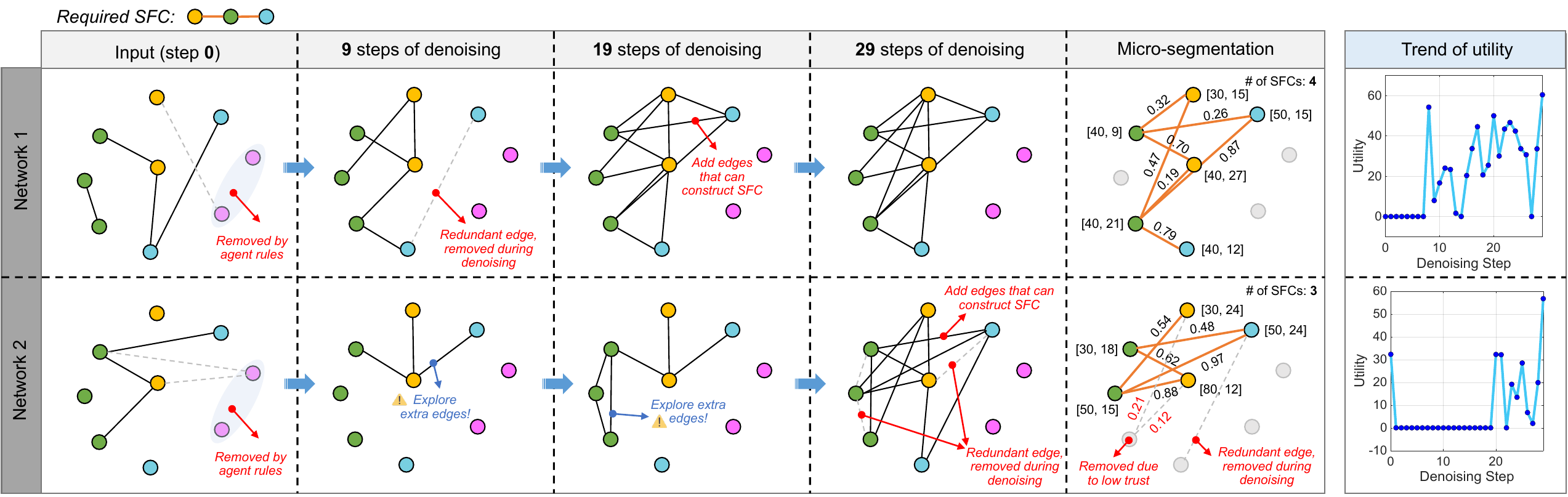} 
  \caption{The case study of micro-segmentation generation. For each node in the generated micro-segmentations, the label [x, y] represents its [computing, transmission] resources. The units are GFLOPs and Watt, respectively.} 
  \label{CASE}
  \vspace{-0.2cm}
\end{figure*}

\subsubsection{Service Provisioning Optimization}
Fig. \ref{TEST1} illustrates the average reward (denoted as \textit{AvgReward}) obtained by LEGD and the aforementioned baselines.
We can observe that no-segmentation acquires an AvgReward of $9.79$ since the entire network is unorganized.
Random segmentation performs better than no segmentation, providing the basic service load balancing.
By greedily segmenting networks via resources and trustworthiness, greedy algorithms can further improve the utility of generated micro-segmentations.
%Compared with the conventional paradigms, our LEGD exhibits great superiority.
With GNNs for representing graphical features and learning policies, DRL greatly outperforms non-learning methods.
%Without enhancement from LLM, it achieves a 39.8\% performance gain than greedy-resource.
The proposed LEGD, however, achieves a higher AvgReward due to the outstanding exploration capability empowered by the generative diffusion architecture.
%Then, we apply the \textit{remove-4} rule generated by the LLM-empowered agent, and the resulting AvgReward increases significantly, achieving a performance gain of over 96\%.
Moreover, we incorporate LLM, which activates the filter that removes all edges connected to type-\textit{4} nodes.
Attributed to the shrunk state space, the performance of LEGD is further improved by 40\%. 
High AvgReward means that the generated micro-segmentation can provision services with high throughput, low latency, and high execution success rate.
Hence, the effectiveness of the proposed LEGD in optimizing service provisioning can be demonstrated.

\subsubsection{Impact of Hyper-parameters}
In this part, we examine LEGD's performance under different configurations and explore the impact of hyper-parameter settings on it. 
First, Fig. \ref{TEST2}(a) illustrates the training curves of LEGD using different learning rates.
We can observe that setting the learning rate to $10^{-4}$ can lead to an AvgReward of \textit{68.5}.
In comparison, training LEGD by a learning rate of $10^{-5}$ achieves faster convergence, while the model fails into sub-optimal.
Learning rate $5\times10^{-5}$ requires more epochs to converge, and the utility of the generated micro-segmentation is lower than that of $10^{-4}$. 
Hence, we fix the learning rate for all the remaining experiments as $10^{-4}$.

Then, Fig. \ref{TEST2}(b) compares the AvgReward of LEGD with varying numbers of diffusion steps.
We can observe performing \textit{45} steps for each denoising process can lead to the best performance.
However, the time consumption of the denoising process increases linearly with the increasing diffusion step numbers \cite{diffusion}.
According to our experiments shown in Fig. \ref{TEST2}(b), generating a 2048-graph batch takes \textit{18.9s}, \textit{37.8s}, and \textit{75.6s}, for \textit{15}, \textit{30}, and \textit{45} diffusion steps, respectively.
Consequently, even if fewer approaches are required to converge, the blue curve's training time is long.
Therefore, we set \textit{30} as the number of diffusion steps, thus striking a balance between performance and training costs.

Finally, we explore the effectiveness of heuristic filter activation by the LLM agent.
%Note that since \textit{remove-4} is validated in Fig. \ref{TEST1}, we only evaluate the trust- and resource-oriented filters in this part.
To do so, we adjust the agent configuration and temperature values to activate different types of filters.
%Note that the meaning of involved filter settings is shown in the caption of Fig. \ref{TEST2}.
As shown in Fig. \ref{TEST2}(c), trust-oriented filters can effectively improve the performance of LEGD since they mask the edges associated with low trustworthiness.
Thus, the denoising network can learn to avoid such edges, ensuring the high execution success rate of the generated micro-segmentations.
Nonetheless, if the heuristic filters remove a large graph part, such as resource-oriented ones and \textit{trust-3}, they may hinder LEGD's exploration of the zero-trust networks, which may affect the performance or even make the model fail into sub-optimal.
In conclusion, incorporating heuristic expertise into graph diffusion can effectively refine models' graphical understanding ability and shrink the state space.
Moreover, the LLM can efficiently generate and activate heuristic filters, thus achieving higher utility for micro-segmentation generations the in most cases.

\subsubsection{Micro-segmentation Generation}
We showcase two denoising processes to illustrate the effectiveness of the proposed LEGD algorithm in controllable micro-segmentation generation.
As illustrated in Fig. \ref{CASE}, the initial micro-segmentation is randomly sampled, containing only a few edges.
With denoising being performed, more edges are generated and involved, thereby forming the final micro-segmentation.
From networks $1$ and $2$, we can observe that LEGD realizes the optimal micro-segmentation generation through the following operations.
First, the heuristic filers generated by the LLM-empowered agent help LEGD remove certain edges, such as the edges connected to type-$\textit{4}$ nodes.
Then, the denoising process will generate new edges.
Since newly generated edges are not guaranteed to make positive progress in optimizing the micro-segmentation, the reward function is utilized to train the denoising network for evaluating each action's desirability.
Undesired edges are removed by the following steps (e.g., step-\textit{0} $\rightarrow$ step-\textit{9} of Network 1).
Finally, the generated edges, as well as the corresponding nodes, form the optimal micro-segmentation.
Moreover, we track utility trends during the denoising process and observe that utility generally increases with more diffusion steps. This indicates the denoising network is well-trained and also validates our reward function. 
\begin{figure}[tbp!]
  \centering
  \includegraphics[width=0.36\textwidth]{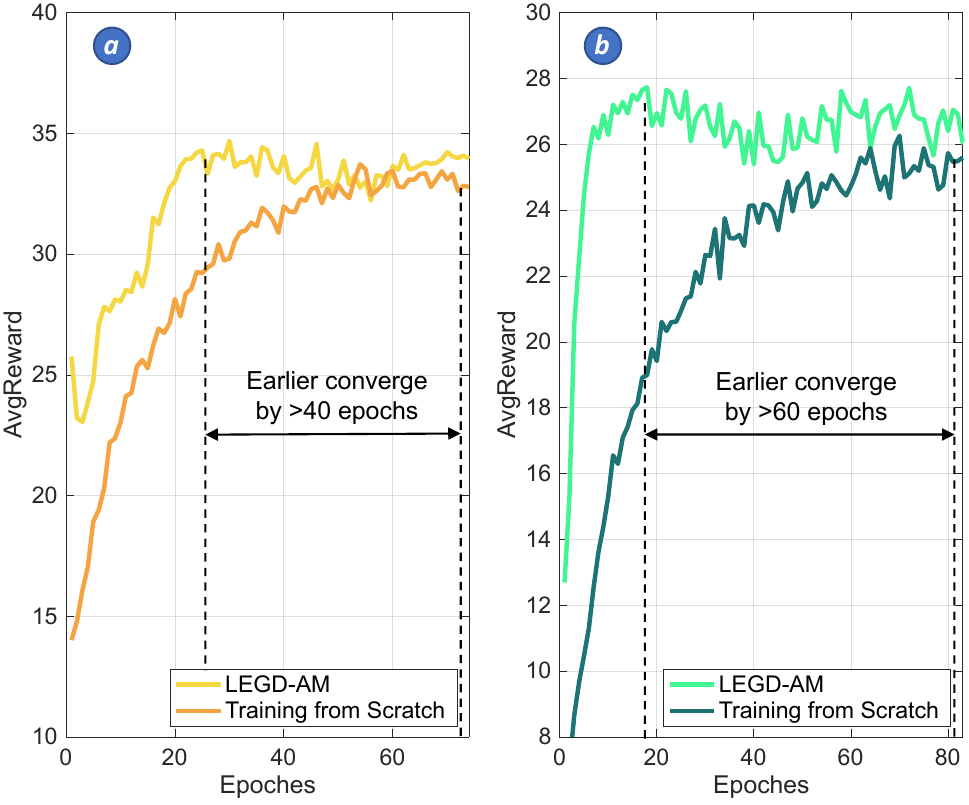} 
  \caption{LEGD-AM and LEGD training curves (i.e., training a model specific for the updated zero-trust NGN). Fig. (a) refers to trustworthiness update, in which a type-\textit{2} node is removed from the current micro-segmentation; Fig. (b) refers to service upgrade, where the \{type-\textit{1}, type-\textit{2}, type-\textit{3}\} service is required to evolve to \{type-\textit{1}, type-\textit{2}, type-\textit{3}, type-\textit{4}\}.} 
  \vspace{0.1cm}
  \label{TEST3}
\end{figure}

\subsection{Efficiency of LEGD-AM}
After the micro-segmentations are generated and operated, they should be maintained continuously to adapt to the dynamic zero-trust NGN.
In this part, we validate the efficiency of the proposed LEGD-AM.

\subsubsection{Fine-tuning on LEGD}
First, Fig. \ref{TEST3} illustrates the train curves of LEGD-AM, where Figs. \ref{TEST3} (a) and (b) correspond to trustworthiness updates and service upgrades, respectively.
To prove the superiority of LEGD-AM, we compare them with training the corresponding LEGD models from scratch.
We can observe that since the well-trained LEGD model for provisioning \{type-\textit{1}, type-\textit{2}, type-\textit{3}\} services is leveraged as the checkpoint, LEGD-AM converges 40 and 60 epochs earlier for trustworthiness updates and service upgrades cases, respectively.
Moreover, it maintains a similar or even better performance than training the model from scratch.
Such experimental results demonstrate the capability of LEGD-AM to promptly adapt to the varying zero-trust NGNs and reduce the service outage time.

\begin{figure}[tbp!]
  \centering
  \includegraphics[width=0.36\textwidth]{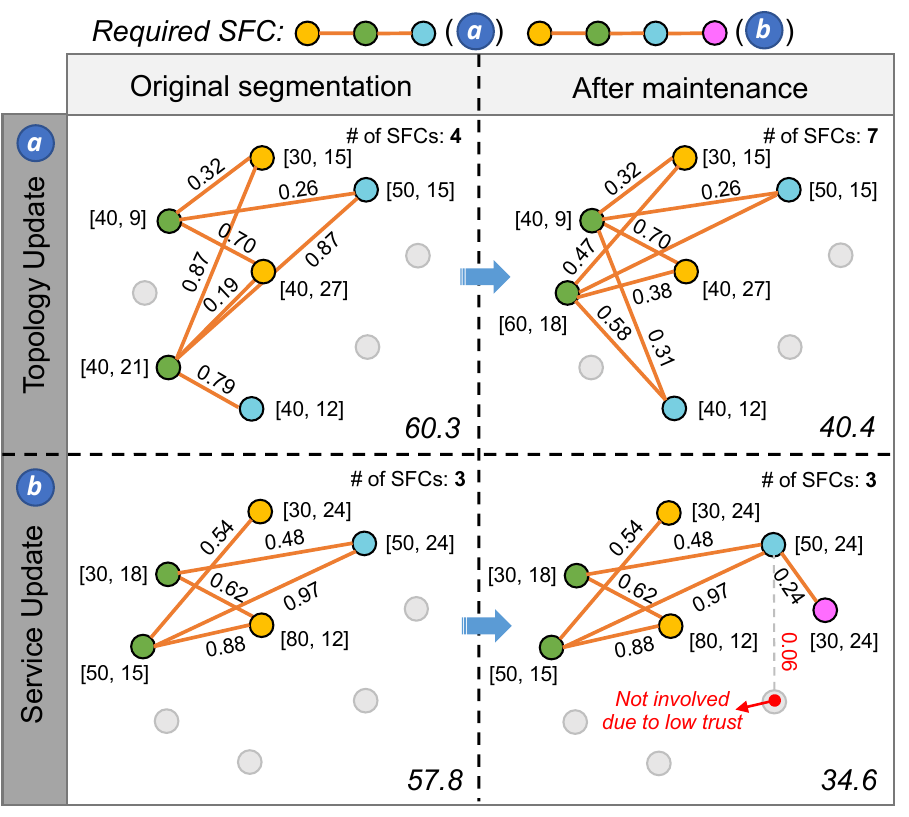} 
  \caption{The case study of adaptive micro-segmentation maintenance. The cases of Figs. (a) and (b) are the same as those in Fig. \ref{TEST3}.} 
  \label{TEST4}
  \vspace{-0.4cm}
\end{figure}

\begin{figure}[tbp!]
  \centering
  \includegraphics[width=0.35\textwidth]{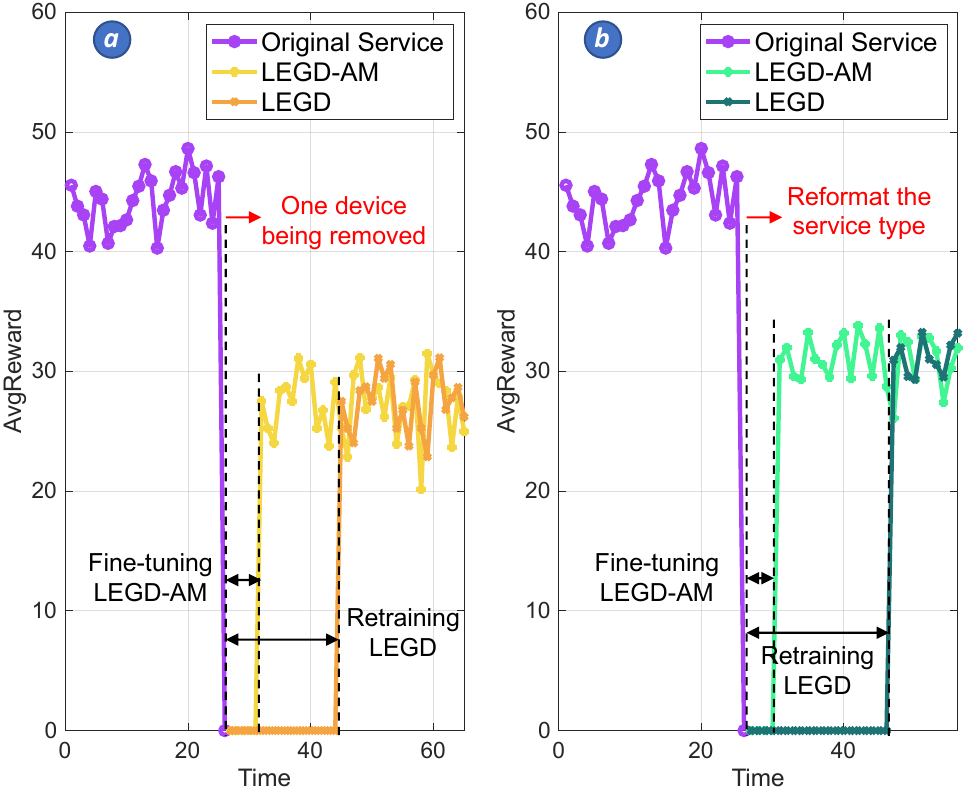} 
  \caption{The operational performance of our proposals. The cases of Figs. (a) and (b) are the same as those in Fig. \ref{TEST3}.} 
  \label{TEST5}
\end{figure}

\subsubsection{Micro-segmentation Maintenance}
Fig. \ref{TEST4}(a) showcases the original and updated micro-segmentation once a type-\textit{2} node is removed.
We can observe that LEGD-AM involves another type-\textit{2} node in the micro-segmentation and re-configures certain SFCs accordingly.
The updated micro-segmentation holds high topological similarity to the original one and also maintains high utility.
Similarly, Fig. \ref{TEST4}(b) illustrates the micro-segmentation for supporting \{type-\textit{1}, type-\textit{2}, type-\textit{3}, type-\textit{4}\} services.
Accordingly, LEGD-AM is trained to refine the micro-segmentation by extending SFCs while minimizing the costs. 
As shown in Fig. \ref{TEST4}, LEGD-AM involves one type-\textit{4} node to the original micro-segmentation, which supports the new service form while minimizing the costs.

\subsection{Operation Performance}
As shown in Figs. \ref{TEST5}(a) and (b), in the beginning, the micro-segmentation generated by well-trained LEGD models can provision user services with high utility.
When network trustworthiness or service states change, we only need to fine-tune the baseline LEGD model following the LEGD-AM paradigm.
Compared with the conventional solutions that re-train the model from scratch, such a strategy can greatly reduce the time of service outage.
In conclusion, the proposed framework and algorithms can effectively address two challenges in service provisioning for zero-trust NGN.

\section{Conclusion}
In this paper, we have presented a systematic method for the efficient provision of zero-trust services in NGN.
Specifically, we have modeled zero-trust networks via hierarchical graphs, leveraging micro-segmentations for network organization and SFCs for service executions.
Based on this framework, we have presented the LEGD algorithm, which leverages graph diffusion, policy optimization, and LLM enhancement to realize the utility-controlled micro-segmentation generation.
Furthermore, we have proposed LEGD-AM, providing an adaptive way to perform task-oriented fine-tuning on LEGD to adapt to new environments/requirements.
Extensive experimental results confirmed the effectiveness of our proposals.

\bibliographystyle{IEEEtran}
\bibliography{zrc_coverage_enhancement}
%\bibliography{mylib}%,IEEEexample}

\end{CJK}
\end{document}